\documentclass[twoside,epsf,11pt]{article}
\usepackage{hyperref}
\usepackage{graphicx} 
\usepackage{multicol} 
\usepackage{floatflt}
 \newcommand{\sun}{{\rm sun}}
 \newcommand{\atm}{{\rm atm}}

\def\CPV{{\begin{picture}(12,0)(0,0)
\put(0,0){\scriptsize CP}\put(0,0){\line(2,1){12}}\end{picture}}}

\def\Red  {}
\def\Black{}
\def\Blue {}

\newcommand{\numextra}{d}
\newcommand{\Wsl}{W\hspace{-9pt}{/}\,\,}
\newcommand{\Zsl}{Z\hspace{-1.4ex}{/}\,}
\def\Red  {}

\def\Black{}
\def\Blue {}
\def\Ord{{\cal O}}
\def\circa#1{\,\raise.3ex\hbox{$#1$\kern-.75em\lower1ex\hbox{$\sim$}}\,}
\newcommand{\eV}{\,{\rm eV}}
\newcommand{\GeV}{\,{\rm GeV}}
\newcommand{\MeV}{\,{\rm MeV}}
\newcommand{\TeV}{\,{\rm TeV}}
\newcommand{\ds}{\partial\!\!\!\raisebox{2pt}[0pt][0pt]{$\scriptstyle/$}}
\newcommand{\Dsl}{D\!\!\!\raisebox{2pt}[0pt][0pt]{$\scriptstyle/$}}
\newcommand{\hc}{\hbox{h.c.}}
\newcommand{\eq}[1]{~{\rm (\ref{eq:#1})}}

\def\Tr{\mathop{\rm Tr}}

\def\One{\hbox{1\kern-.24em I}}
\newcommand{\NP}{Nucl. Phys.}
\newcommand{\JHEP}{J.HEP}
\newcommand{\PRL}{Phys. Rev. Lett.}
\newcommand{\PL}{Phys. Lett.}
\newcommand{\PR}{Phys. Rev.}
\makeatletter
\def\art{\@ifnextchar[{\eart}{\oart}}
\def\eart[#1]#2#3#4#5#6{{\rm #2}, {\em #3 \rm #4} {\rm (#6) #5} ({\em #1})}
\def\hepart[#1]#2{{\rm #2, \em#1}}
\newcommand{\oart}[5]{{\rm #1}, {\em #2 \rm #3} {\rm (#5) #4}}

\begin{document}
%\twocolumn[
\centerline{\today \hfill    CERN-TH/2001--184}
\centerline{hep-ph/0107156 \hfill IFUP--TH/18--2001} \vspace{1cm}
\begin{center}
{\LARGE\bf\Red Phenomenological implications of \\
neutrinos in extra dimensions }\Black
\end{center}
\bigskip
\begin{center}
{\bf Andr\'e de Gouv\^ea, Gian Francesco Giudice,\\
Alessandro Strumia\footnote{On leave from
 Dipartimento di Fisica dell'Universit\`a di Pisa and INFN, Italy.}, and Kazuhiro Tobe}
\end{center}
\bigskip
\centerline{\em Theoretical Physics Division, 
CERN, CH-1211, Gen\`eve 23, Switzerland}
\bigskip\bigskip

\centerline{\large\bf\Blue Abstract}
\begin{quote}\indent\large
Standard Model singlet neutrinos propagating in extra dimensions induce small 
Dirac neutrino masses. While it seems rather unlikely that their
Kaluza-Klein excitations directly participate in the observed neutrino 
oscillations, their virtual exchange may lead to detectable 
signatures in future neutrino experiments and in rare charged lepton processes.
We show how these effects can be described by specific dimension-six effective 
operators and discuss their experimental signals.\Black
\end{quote}

\vspace{1cm}

\setcounter{equation}{0}
\section{Introduction}

The hypothesis that Standard Model (SM) singlet fields propagate in  
extra dimensions leads to striking results.
When applied to the graviton, it allows 
to lower the quantum gravity scale down to few TeV~\cite{TeVgravity,RS},
suggesting a new scenario for addressing the Higgs mass hierarchy problem.
It is also natural to consider the case of 
``right-handed neutrinos'' ({\it i.e.,}\/ 
fermions without SM gauge interactions) 
propagating in extra dimensions.
The smallness of the neutrino masses, of the Dirac type, could in fact be a 
manifestation of this hypothesis~\cite{nuR5dA,DDG,nuR5d,GN}. 

If the radius of the compactified dimensions is very large, $R\circa{>} 
\eV^{-1}$, Kaluza-Klein (KK) modes of right-handed neutrinos would significantly
participate in  neutrino oscillations. However, KK interpretations of
the atmospheric and solar neutrino puzzles are disfavoured 
by the following arguments:

\begin{itemize}

\item A KK tower of sterile neutrinos gives rise to active/sterile oscillations at
a small $\Delta m^2\sim 1/R^2$ only in the case of a 
{\sl single large}\/ extra dimension. In the case of more large extra dimensions, 
the active/sterile mixing is not dominated by the lightest KK modes, and the 
mixing with the heaviest KK modes does not lead to oscillations. One could argue
that right-handed neutrinos in a single very large extra dimension can be effectively
obtained if one of the extra dimensions happens to be much larger than the other(s). 
However, the fact that infrared effects are dominant in this case
destabilises the hierarchy~\cite{Antoniadis}:
the Newton or Coulomb potential grows with the size $R$ of the largest dimension.
%%  ($V\propto R^{2-\delta}$, where $\delta$ is the number of largest extra dimensions).

\item Even if one forgets about the hierarchy problem, 
%%and considers  the models with right-handed neutrinos living in one large extra  
%%dimension, 
there are severe bounds from supernov\ae{} observations~\cite{amen}.  
One needs to prevent resonant neutrino conversion in supernov\ae{} by choosing a 
small radius $R\circa{<}1/\MeV$ or by adding an ad-hoc 5-dimensional mass term 
$m\circa{>} \MeV$  for the right-handed neutrino(s) responsible for 
``atmospheric'' oscillations. 
The latter still allows to build models for the solar 
anomaly roughly compatible with supernov\ae{} bounds~\cite{AccanimentoTerapeutico}.

\item 
Finally,
the recent results from the SuperKamiokande~\cite{SKatm,SKsun} and SNO~\cite{SNO} 
experiments provide 
strong indications for $\nu_\tau$ appearance in atmospheric oscillations and for
$\nu_\mu,\nu_\tau$ appearance in solar oscillations.
In both cases, a sterile interpretation is now strongly disfavoured by experiments.

\end{itemize}
In this paper we explore the phenomenology of more promising models
with $\delta > 2$ extra dimensions and radii which are not larger than
what is required to reproduce the gauge/gravitation hierarchy.
The neutrino puzzles are solved by ``normal'' (active) oscillations,
but the presence of the {\sl heaviest}\/ KK neutrinos can still lead to small 
but detectable effects in neutrino flavour transitions.
After integrating out the heavy KK modes, we obtain an effective Lagrangian that
contains massive Dirac neutrino states and a specific set of non-renormalizable 
operators. Since the dominant effects come from the heaviest KK states, the 
coefficients of these operators can only be estimated by introducing an arbitrary 
ultraviolet cut-off.

At tree level one only obtains the dimension-six operators 
\begin{equation}\label{eq:TreeLevel}
%%\frac{\epsilon_{ij}}{v^2} (H^\dagger \bar{L}_i)i\ds (HL_j),
{\cal L}_{\rm tree} =  \epsilon_{ij}~
2\sqrt{2}G_F (H^\dagger \bar{L}_i)i\ds (HL_j),
\end{equation}
where $L_i$ are the lepton left-handed doublets, $H$ is the Higgs doublet, 
%%$v$ is its vev, 
and $\epsilon_{ij}=\epsilon_{ji}^*$ are dimensionless couplings.
%It conserves total lepton number and affects only neutrino physics.
The presence of these effective operators leads, for example, 
to potentially large flavour transitions $P(\nu_i\to
\nu_j)\sim |\epsilon_{ij}^2|$  at ${\cal O}(L^0)$ ({\it i.e.,}\/ at very short
baselines $L\ll E_{\nu}/\Delta m^2$),
%in short baseline neutrino experiments ($L\ll E_{\nu}/\Delta m^2$),
%~(\ref{eq:TreeLevel}) induces potentially large $P(\nu_i\to
%\nu_j)\sim |\epsilon_{ij}^2|$  at ${\cal O}(L^0)$, 
and CP-violating effects at ${\cal O}(L^1)$ 
(rather than at ${\cal O}(L^2)$ and ${\cal O}(L^3)$ as in ordinary oscillations). 
Since this peculiar tree level operator {\sl only affects 
neutrinos}\/, potentially detectable effects, especially at a neutrino 
factory, 
are not already excluded by bounds from rare charged leptons processes, like
$\tau\to \mu\gamma$, $\mu\to e \gamma$, $\mu\to e\bar{e} e$, etc.
We will show, however, that Eq.\eq{TreeLevel} cannot explain the LSND 
anomaly~\cite{LSND}, due to the present constraints from searches for
$\mu\rightarrow e\gamma$.

At one-loop level, other operators are 
generated, giving rise to rare muon and tau processes that violate 
lepton flavour. Again, the coefficients of these operators  are cut-off dependent 
and can only be estimated. However, in minimal models, their flavour structure is
directly related to the physical neutrino mass matrix, giving 
predictions for rare muon processes
in terms of neutrino oscillation parameters. This is in contrast with
other cases of physics beyond the SM. In supersymmetry, for instance,
the rates for rare muon processes are perturbatively calculable, but their
relations with neutrino oscillations parameters are strongly model-dependent
and can vary by many orders of magnitude. 

Many of the effects studied here have already been considered in previous 
analyses~\cite{Faraggi,Pilaftsis} as due to mixing between ordinary 
neutrinos and 
the whole tower of KK 
states. The equivalent language of effective operators we employ
allows the study of different models ({\it e.g.}\/ large and warped extra dimensions) 
in a unified framework, the discrimination of what is really computable
from what can only be estimated, and a more transparent 
identification of all potentially interesting experimental signatures.

The paper is organized as follows. In Sec.~2, we describe the models 
under consideration. 
In Sec.~3 we discuss the effects of the tree level operator Eq.~(\ref{eq:TreeLevel}) 
in neutrino physics. In Sec.~4 we discuss the effects of the operators induced 
at one-loop in charged lepton processes. In Sec.~5 we summarize our results.
%(it may prove useful to read it before getting lost with details).

\setcounter{equation}{0}
\setcounter{footnote}{0}
\section{Right-handed neutrinos in extra dimensions}
We will study models with large flat extra 
dimensions~\cite{TeVgravity,nuR5dA,DDG,nuR5d} and models 
with one warped extra dimension~\cite{RS,GN}.

\subsubsection*{Large flat extra dimensions}
We consider  $(4+\delta)$-dimensional massless fermions 
$\Psi_i(x_\mu,y)$ which, inside their $2^{(4+\delta)/2}$ components
(for even $\delta$) or $2^{(3+\delta)/2}$ components (for odd $\delta$),
contain the degrees of freedom of the ``right-handed neutrinos'' $\nu_{Ri}$
($i=1,2,3$ is the generation index). The fermions $\Psi_i$
interact in our brane, through their components $\nu_{Ri}$,  
with the standard left-handed lepton doublet $L_i$ in a way that conserves total 
lepton number. The relevant part of the action is
\begin{equation}
\label{eq:5action}
{\cal S}=
\int d^4\!x~d^\delta\! y~ \bar{\Psi}_i i\Dsl\, \Psi_i + 
\int d^4\! x~\left[ \bar{L}_i i\ds \,L_i - \left(
\bar{L}_i \lambda_{ij} \nu_{Rj}
H^\dagger   +\hc \right) \right],
\end{equation}
where $i\Dsl$ is a $4+\delta$ dimensional Dirac operator,
$H$ is the SM Higgs doublet in four 
dimensions and $\lambda$ is a matrix
of Yukawa couplings with dimensions ${\rm (mass)}^{-\delta/2}$. 
As is manifest from Eq.~(\ref{eq:5action}), $\lambda$ can be made 
diagonal without loss of generality at the price of introducing the usual 
unitary matrix $U$, which describes flavour-changing charged current 
neutrino interactions. 
% The matrix $U$ contains 3 mixing angles and one CP violating  phase. 

Using the KK decomposition
\begin{equation}
\Psi_i(x,\vec{y})= \frac{1}{\sqrt{V_\delta}}\sum_{\vec{n}} \Psi_{\vec{n}i}(x) \exp
\left(\frac{i \vec{n}\cdot\vec{y}}{R}\right) ,
\end{equation}
where $V_\delta \equiv (2\pi R)^\delta$ is the compactified volume, and
performing the $d^\delta y$ integration, Eq.~(\ref{eq:5action}) yields the four dimensional
neutrino Lagrangian
\begin{equation}
{\cal L}= \bar{L}_i i\ds \,L_i +\sum_{\vec{n}} \left\{\bar{\Psi}_{\vec{n}i} \left(i\ds\,-
\frac{\vec{n}\cdot \vec{\gamma}}{R}\right) \Psi_{\vec{n}i} - 
\left[ \frac{\lambda_{ij}}{\sqrt{V_\delta}}
 \bar{L}_i \nu_{R\vec{n}j}  H^\dagger +\hc \right] \right\},
\label{lagrquat}
\end{equation}
where $\vec{\gamma}$ are the extra-dimensional Dirac matrices.
After electroweak symmetry breaking, the neutrinos 
obtain a Dirac mass matrix
$m_\nu = \lambda v/\sqrt{V_\delta}$,
where $v=174\GeV$ is the vacuum expectation value of the Higgs boson.
The heavy modes give corrections suppressed by $R m_\nu$,
that are negligible in the cases of interest.

\medskip

In order to compute the tree-level effects due to the presence of the tower of KK 
states, we construct an effective Lagrangian by substituting in Eq.~(\ref{lagrquat})
the solutions of the equations of motion for the heavy fields,
\begin{equation}
 \Psi_{\vec{n}i} = \left(i\ds\,-
\frac{\vec{n}\cdot \vec{\gamma}}{R}\right)^{-1}
\frac{\lambda^*_{ij}}{\sqrt{V_{\delta}}} L_j H .  
\label{inver}
\end{equation}
With an abuse of notation, we have identified $L_j$ in Eq.~(\ref{inver})
with a higher-dimensional spinor,
%%  with a $2^{(4+\delta)/2}$  (for even $\delta$) or $2^{(3+\delta)/2}$ (for odd $\delta$) dimensional  vector,
in which $L_j$ fills the components
corresponding to the right-handed neutrinos, while all other components are zero.
Ignoring higher derivative terms and summing over all KK states up to an ultraviolet 
cut-off $\Lambda$,
we obtain the effective operator in Eq.\eq{TreeLevel} with coefficients
\begin{equation}
\epsilon_{ij}= 
   \ell_{\delta} \frac{(\lambda \lambda^\dagger)_{ij}
   \Lambda^{\delta-2} v^2 }{\delta-2}
 ,\qquad
   \ell_\delta\equiv \frac{S_\delta}{(2\pi)^\delta},
   \label{pitpit}
\end{equation}
%   \begin{equation}
%   {\cal L}_{\rm tree} = 
%   %\frac{2\pi^{\delta/2}}{\Gamma(\delta/2)}
%   \ell_{\delta} \frac{(\lambda \lambda^\dagger)_{ij}
%   \Lambda^{\delta-2}  }{\delta-2}
%   (H^\dagger \bar{L}_i)i\ds (HL_j) ,\qquad
%   \ell_\delta\equiv \frac{S_\delta}{(2\pi)^\delta}
%   \label{pitpit}
%   \end{equation}
where $S_\delta={2\pi^{\delta/2} / \Gamma(\delta/2)} $ is the surface of a
unit-radius $\delta$-dimensional sphere, 
$\lambda$ is the dimensionful matrix of
Yukawa couplings and $\Lambda$ parameterizes 
some ultraviolet cut-off. 
We take  $\Lambda$ to be the KK mass at which we cut off
the summation.  The operator in Eq.~(\ref{eq:TreeLevel})
is gauge invariant, in
spite of containing an ordinary derivative (rather
than a covariant one), because $HL$ is a SM gauge singlet. The coefficients
$\epsilon_{ij}$ of this 
operator have the same flavour structure as the physical neutrino masses
$m_\nu m_\nu^\dagger \propto \lambda\lambda^\dagger$, although its overall
factor is model-dependent. The order of magnitude of this single free parameter 
is fixed if --- motivated by the hierarchy problem --- we assume that all 
extra-dimensional physics is at the TeV scale, 
$\lambda^{-2/\delta}\sim \Lambda\sim \TeV$. 
In this case the coefficients $\epsilon_{ij}$  in Eq.~(\ref{pitpit})
are of order $\ell_\delta (v/\TeV)^2$,
where $\ell_\delta$ is a typical loop factor in  $\delta$ dimensions.
It is useful to write the coefficient of Eq.~(\ref{pitpit}) as
$\epsilon_{ij} \equiv \epsilon (m_\nu m_\nu^\dagger)_{ij}/\Delta m^2_{\atm}$,
because $\epsilon$ is the only free parameter of the model (at 
tree level).\footnote{Throughout the
paper, we will assume that the neutrino masses obey a normal hierarchy, {\it i.e.,}\/
$m_1^2\ll m_2^2 \ll m_3^2$, so that $\Delta m^2_{\rm atm}\approx m_3^2$ and 
$\Delta m^2_{\sun}\approx m_2^2$.}
Explicitly,
\begin{equation} 
\epsilon = \frac{\ell_{\delta}}{\delta-2}\frac{\Delta m^2_{\atm}}{\Lambda^2}
V_{\delta}\Lambda^{\delta}.
\label{eq:def_epsilon}
\end{equation}

We now recall how this model can be linked with the Higgs mass hierarchy 
problem.
\begin{floatingfigure}[r]{10cm}
$$\includegraphics[width=10cm]{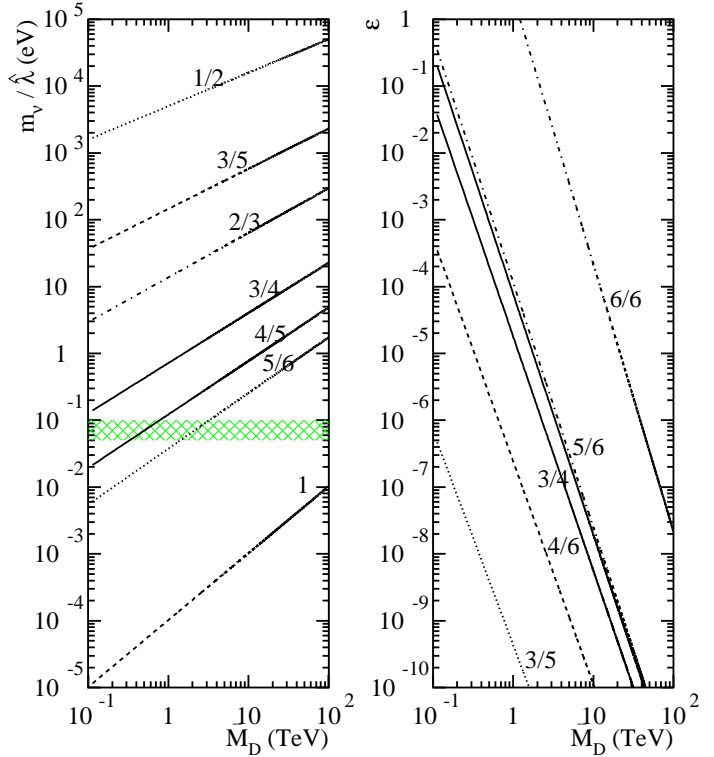} $$
\caption{\em $m_\nu /\hat \lambda$ (left) and $\epsilon$ (right) 
as a function of $\bar{M}_{D}$ for different values of
$\delta/\numextra$, assuming
$\Lambda=\bar{M}_{D}$. The horizontal band in the left panel shows the mass
range selected by atmospheric neutrino data for $\hat \lambda =1$.}\vspace{8mm}
\label{fig:md_and_epsilon}
\end{floatingfigure}
If gravity propagates in $\numextra$ extra dimensions ($\numextra \ge \delta$,
where all extra dimensions have the same radius $R$ and the topology of a
torus), the reduced Planck mass ${\bar M}_4= 2.4\times 10^{18}\GeV$ 
is related to the reduced Planck mass $\bar{M}_D$ in $D=4+\numextra$ dimensions by
  \begin{equation}
  {\bar M}_4^2=(2\pi R)^\numextra \bar{M}_D^{2+\numextra}.
 \label{Mp}
  \end{equation}
It is useful to rewrite the dimensionful Yuka\-wa coupling $\lambda$ in terms of a 
dimensionless Yukawa parameter $\hat{\lambda}$ as
\begin{equation}
\lambda \equiv \hat{\lambda}/\bar{M}_D^{\delta/2}.
\end{equation}
With this definition, the neutrino mass becomes
\begin{equation}
m_\nu =\hat{\lambda} v \left( \frac{\bar{M}_D}{{\bar M}_4}\right)^{\delta /
\numextra}.
\label{nema}
\end{equation}
The simplest case $\delta=\numextra$ requires
very large $\hat{\lambda}$ or $\bar{M}_D$ in order to obtain neutrino masses 
that satisfy the atmospheric neutrino data (see 
Fig.~\ref{fig:md_and_epsilon}).
On the other hand, for $\delta =5$ and $\numextra =6$, 
\begin{equation}
m_\nu = \hat{\lambda} \left( \frac{\bar{M}_D}{\rm TeV}\right)^{5/6} ~3
\times 10^{-2}~{\rm eV},
\end{equation}
and the atmospheric mass scale is obtained with $\hat{\lambda}\sim 1$ and 
$\bar{M}_D\sim \TeV$, while the solar neutrino puzzle can be solved for values of 
$\hat{\lambda}$ slightly smaller than one. Figure~\ref{fig:md_and_epsilon} (left) 
depicts values
of $m_\nu/\hat{\lambda}$ as a function of $\bar{M}_{D}$, for different values of 
$\delta/\numextra$.
Supernov\ae{} bounds~\cite{amen} force all KK neutrino states to be heavier than the 
typical supernova temperature $T\sim 30\MeV$. In the present model, these bounds 
imply $\numextra \circa{>}3$. The reduced Planck mass $\bar{M}_D$ is related
to the phenomenological parameter $M_D$ used to study graviton effects at colliders as
$M_D = (2\pi)^{\numextra/(2+\numextra)} \bar{M}_D$.
The present collider bound on $\bar{M}_D$ is $\bar{M}_D\circa{>}200\GeV$ (for 
$\numextra\sim 6$) while the LHC can improve it by a 
factor $\circa{<}10$~\cite{ColliderGravitonico}.
Fig.~\ref{fig:md_and_epsilon} (right) depicts the values of
$\epsilon$, defined in Eq.~(\ref{eq:def_epsilon}),
as a function of $\bar{M}_D$ in the case $\Lambda=\bar{M}_D$
for different values of $\delta$ and $\numextra$, and for $\Delta m^2_{\atm}=
3\times 10^{-3}$~eV$^2$. It is important to keep in mind that
$\epsilon$ is strongly enhanced if $\Lambda>\bar{M}_D$ (it scales like 
$(\Lambda/\bar{M}_D)^{\delta}$). For a fixed ratio 
$\Lambda/\bar{M}_D$, $\epsilon$ decouples like $1/\Lambda^2$, as the
new-physics scale increases. This behavior is clearly visible in 
Fig.~\ref{fig:md_and_epsilon}.

One may consider a few variations to the ``minimal'' model described above:
\begin{itemize}
\item[(A)] The right-handed neutrinos could have some extra dimensional mass term 
$m \circa{>} m_\nu$~\cite{DDG}. The mass term $m$ does not affect the dimension-six 
operators, but affects neutrino masses --- now related to $m$ and $\lambda$ by
a higher-dimensional ``see-saw'' relation, $m_\nu \propto \lambda^2/m$. 
In these models, the $\epsilon_{ij}$ coefficients 
are not directly related to neutrino 
masses, and therefore contain additional mixing angles and CP violating phases
beyond the ones in the neutrino mixing matrix.\footnote{If $m\sim m_\nu$, 
more than three mass eigenstates participate in oscillations.
Explaining the origin of a higher-dimensional mass term comparable to the neutrino masses
is, perhaps, the biggest challenge for this type of scenario.}

\item[(B)] Different massless `right-handed neutrinos' could have a 
different UV cut-off, or live in a different number of extra dimensions. 
An interesting case is obtained with 2 right-handed neutrinos 
living, respectively, in 5 and 6 extra dimensions with equal radii.
In this case one can reproduce the smallness of the solar $\Delta m^2$ with respect 
to the atmospheric $\Delta m^2$ using comparable Yukawa couplings (see 
Fig.~\ref{fig:md_and_epsilon}). 
%% Note that in this case one of the neutrinos is  massless. 
%   $\delta_{\atm} = 5$ and $\delta_{\sun}=6$:
%   \begin{equation}\label{eq:action2}
%   S=
%   \int d^4\!x~d^{\delta_{\atm}}\! y~[\bar{\Psi}_1 i\ds\, \Psi_1] + 
%   \int d^4\!x~d^{\delta_{\sun}}\! y~[\bar{\Psi}_2 i\ds\, \Psi_2] + 
%   \int d^4\!x~[\bar{L}_i i\ds \,L_i + {\bar L}_i \lambda_{ij}P_R\Psi_j 
%   H^\dagger \delta^\delta (y)  +\hc]
%   \end{equation}
This model contains 4 mixing angles and 2 CP-violating phases
(rather than the 3 mixing angles and 1 CP phase of the minimal model).
The reason is that, unlike in the minimal model, it is not possible to perform flavour 
rotations of the
right-handed neutrinos.

\end{itemize}
Qualitatively, the main new feature of non minimal models
is that the tree level exchange of bulk neutrinos again yields the operator 
Eq.\eq{TreeLevel},
but the overall coefficient is now different for the ``atmospheric'' and ``solar'' 
contributions.
Assuming that all Yukawa couplings are of order of the ``fundamental'' mass scale,
in case (B) the ``solar'' contribution becomes comparable to the ``atmospheric'' 
contribution,
%{\it i.e.,}\/ $\epsilon_{e \mu}\sim \epsilon_{\mu \tau}$ (for large solar mixing).
{\it i.e.,} all $\epsilon_{ij}$ in Eq.~(\ref{eq:TreeLevel})
are expected to be of the same order, contrary to the 
minimal model where $\epsilon_{\mu\tau}/\epsilon_{e\mu}\propto 
\Delta m^2_{\atm}/\Delta m^2_{\sun}$, assuming
that $|U_{e3}|$ is negligible (this will be discussed in detail in the next 
section). 
Warped models generically give effects qualitatively similar to case (B), as we 
describe below.

\subsubsection*{Warped extra dimensions}
A strong suppression of gravity could
be generated by an extra-dimensional red-shift factor~\cite{Gog,RS}.
We will concentrate on the simplest scenario~\cite{RS}, containing one extra dimension
with topology $S_1/Z_2$ ({\it i.e.,}\/ a segment), parameterized by a coordinate $y$.
The SM fields live on a brane fixed at one of its borders ($y=R$),
and another brane lives at the opposite border ($y=0$).
After an appropriate fine-tuning of the cosmological constant $\Lambda$ in the 
extra dimension and the tension of the two branes, the background metric has 
the form $ds^2 = e^{-2ky} \eta_{\mu\nu} dx^\mu dx^\nu +  dy^2$,
where the warping constant $k$ is determined by the five-dimensional cosmological 
constant $\Lambda$ and
gravitational scale $M$ as
$k^2 = -\Lambda/24M^3$ \cite{RS}.
By rescaling the metrics at $y=0$ to its canonical form, one finds that all mass scales
in the four-dimensional SM Lagrangian
get red-shifted by a factor $e^{-kR}$
(including the ones that should suppress unwanted non renormalizable operators),
and that the four-dimensional Planck mass is $M_{\rm Pl}^2 = (1-e^{-2kR})M^3/k $.
One can try to explain the gauge/gravitation hierarchy by
stabilizing the size of the extra dimensions $R$~\cite{GW}
such that $e^{-kR}\sim \TeV/M_{\rm Pl}$.\footnote{It has been conjectured that this 
model is equivalent to walking {\sl composite} technicolour~\cite{Witten}. 
This dual version is not as \ae{}stetichally appealing,
but its problems with experimental data are better known.}

It is again interesting to consider five-dimensional `right-handed neutrinos'
with Yukawa couplings $\lambda$ to the SM fields living on a brane.
In this case it is {\sl necessary} to give some five-dimensional Dirac mass terms 
$m$ to the extra-dimensional
neutrinos in order to naturally explain the smallness of the neutrino 
masses~\cite{GN}. Despite this higher dimensional mass term, there is still 
one very light KK mode, and its effective Yukawa coupling with the active neutrinos is
strongly suppressed if $m>k/2$. Then, the Dirac neutrino masses
depend very strongly on $m$,  $m_\nu\sim v e^{-mR}$,
and neutrino masses which span many orders of magnitude are
easily obtained even if all the higher dimensional Yukawa couplings are comparable.

The other KK states have TeV-scale masses and unsuppressed four-dimensional 
Yukawa couplings to SM fields. 
They generate the operator Eq.\eq{TreeLevel} with 
coefficient
\begin{equation}
\epsilon_{ij} \sim \frac{e^{2kR} v^2}{k} (\lambda\lambda^\dagger)_{ij}.
\end{equation}
Because this is a 4+1 dimensional theory, the infinite sum over all KK modes is finite,
and the introduction of an arbitrary ultraviolet cut-off is unnecessary.
On the other hand, it is not useful to perform a precise computation because,
similarly to the non minimal flat models, there is no  
direct connection between  $\epsilon_{ij}$ and the neutrino masses.
The main conclusion is that all $\epsilon_{ij}$ have comparable values,
$\epsilon_{ij}\sim 10^{-{\rm few}}$, if $\lambda$, $k$, 
$M$ have comparable mass scales. 
Smaller $\epsilon_{ij}$ can be obtained for smaller Yukawa couplings $\lambda$.

We do not consider ``non minimal'' warped models, because they do not seem to lead to new 
interesting effects in neutrino physics.\footnote{In more general extra
dimensional black-hole backgrounds, 
the space warping factor can be different from the time
warping factor~\cite{etere}. This could lead to right-handed neutrinos that travel 
with a velocity $c(1+\delta c)$ different from light.
This model has a conjectured holographic dual~\cite{Witten} where
Lorentz invariance is broken in a more obvious and old way~\cite{etereold}:
normal right-handed neutrinos have an index of refraction $n$ due to interactions
with an `\ae{}ther' (composed of some hot conformal matter).
At tree level and up to negligible $m/E$ corrections,
these effects do not affect neutrino oscillations
as long as there are only Dirac mass terms~\cite{MSW}, 
or more generically as long as the
energy eigenstates do not mix left with right-handed neutrinos.
Even in models where this not the case, neutrino effects caused by a
non universality of the speed of the light
do not seem phenomenologically interesting because
have an energy dependence (different from the observed one~\cite{n=1}) that allows
to derive very strong constrains.
Upward through going atmospheric muons in SuperKamiokande (with $L\sim 10^4\,{\rm km}$ 
and $E_\nu \sim \TeV$) put the bound $\delta c \circa{<} 1/L E_\nu\sim 10^{-(24\div 25)}$ (for 
large mixing between flavour and velocity eigenstates), preventing detectable effects 
in planned new neutrino experiments.}
%% and in Michelson-Morley experiments

\setcounter{equation}{0}
\setcounter{footnote}{0}
\section{Tree level effects in neutrinos}

The tree-level operator Eq.~(\ref{eq:TreeLevel}) only affects 
neutrinos.\footnote{The operator in Eq.~(\ref{eq:TreeLevel}) 
does not contribute to the $h\to
\nu\bar{\nu}$ decay, due to conservation of angular momentum.
Only the emission of KK states lighter than $m_H$ yields $h\to \nu \bar\nu_n$ 
corrections to 
Higgs decays~\cite{nuR5d,Agashe}.} For this reason, Eq.~({\ref{eq:TreeLevel}) 
can give rise to detectable effects in neutrinos which are 
compatible with bounds from charged-lepton 
processes, affected only at the one loop order. After electroweak symmetry breaking, 
the operator in Eq.~(\ref{eq:TreeLevel}) contributes to the kinetic term of the 
neutrinos. The effective Lagrangian for the 3 left-handed neutrinos $\nu$ and for the
zero mode right-handed neutrinos $\nu_R$  becomes
\begin{eqnarray}
{\cal L}&=&\bar{\nu}_i i\ds  (\delta_{ij} + \epsilon_{ij}) \nu_j +
\bar{\nu}_R i\ds   \nu_R 
-\left( \bar{\nu} m_\nu  \nu_R + \hc \right) 
+\left( \frac{g}{\sqrt{2}}\bar{\ell}\Wsl\nu  +\hc \right)
+ \frac{g}{2\cos \theta_{\rm W}}\bar{\nu} \Zsl   \nu . 
\label{eq:def_a}
\end{eqnarray}
where $\ell = (e_L,\mu_L,\tau_L)$ are the left-handed  charged leptons
in the mass eigenbasis, 
$m_\nu$ is the neutrino mass matrix, 
%of Eq.~(\ref{nema})
and $\epsilon_{ij}$ are adimensional
numbers defined in Eq.~(\ref{pitpit}), expected to be of order    
$\ell_\delta (v/\TeV)^2$.
The kinetic and mass terms can be simultaneously diagonalized by the following 
{\sl non-unitary}
field redefinitions
\begin{equation}
\nu_R \to U_R \nu_R ,~~~~\nu_L \to U_L K U'_L \nu_L.
\end{equation}
Here $U$ are unitary matrices (such that $U_L$ diagonalizes $\epsilon_{ij}$) 
and $K$ is a diagonal matrix whose elements are
$1/\sqrt{1+\epsilon_i}$, where $\epsilon_i$ are the eigenvalues of $\epsilon_{ij}$.
In this new basis the neutrino kinetic and mass terms are flavour diagonal, but
the neutrino interactions with the gauge bosons are modified
\begin{equation}
{\cal L}=\bar{\nu}  i\ds  \nu +
\bar{\nu}_R i\ds   \nu_R -
(\bar{\nu}_n m_n \nu_{Rn} +\hc)
+\left( \frac{g}{\sqrt{2}}\bar{\ell } V_W \Wsl \nu +\hc \right)
+ \frac{g}{2\cos \theta_{\rm W}}\bar{\nu} V_Z  \Zsl   \nu , 
\end{equation}
where $m_n$ are the mass eigenvalues given by $m_{\nu n}/\sqrt{1+\epsilon_n}$, 
$V_W\equiv U_L K U'_L $ and 
$V_Z \equiv U^{\prime\dagger}_L K^2 U'_L$.
It is now easy to compute oscillation probabilities. 
The transition probability for a neutrino both produced and detected 
via a charged-current $W$ interaction is given by
\begin{equation} \label{eq:PWW}
P(\nu_i\to \nu_f) =  \left|\sum_n V^W_{i n} V^{W*}_{f n} 
e^{i E_n t } \right|^2, \qquad E_n - E_m =  \frac{\Delta m^2_{nm}}{2E}. 
\end{equation}
Differently from the standard case, this expression  cannot be simplified to the usual 
form, because $V^W$ is not a unitary matrix.

The minimal model described in detail in the previous section predicts
$\epsilon_{ij} \propto (m_\nu m_\nu^\dagger)_{ij}$,
such that $U_L'=\One$.
The transition  probability reduces to the more transparent form
\begin{equation}
P(\nu_i\to \nu_f) = \left|\sum_n \frac{U_{i n} U_{f n}^*}{1+ 
\epsilon m_n^2/\Delta m_{\atm}^2} e^{i E_n t } \right|^2,
\end{equation}
where $U$ is the usual unitary neutrino mixing matrix.
Notice that the summed survival probability $\sum_{f=e,\mu ,\tau}  
P(\nu_i\to \nu_f)
\simeq 1-2 \epsilon$ is not equal to 1, because of the non-unitary
$\epsilon$ effects. This corresponds to a non-vanishing transition 
probability into sterile KK modes or, in the effective theory, 
to a reduced interaction of the neutrinos with the $W$ boson.

%The dimensionless parameter $\epsilon$ was introduced in the previous section 
%[see Fig.~\ref{fig:md_and_epsilon}(right)].
%Fig.~\ref{fig:md_and_epsilon}(right) depicts values of $\epsilon$ as a function of 
%$\bar{M}_{D}$ for different values of $\delta$ and $\numextra$, in the case $\Lambda = 
%\bar{M}_{D}$. Note that $\epsilon$ scales like $(\Lambda/\bar{M}_{D})^{\delta}$. 

Next, we discuss a few possible experimental signals, concentrating on searches for
neutrino oscillations.

\subsubsection*{Flavour transitions at very short baselines}
In terms of the $\epsilon_{ij}$ parameters in Eq.\eq{TreeLevel},
the transition probability at a very short baseline ($L\approx 0$)
derived from Eq.~(\ref{eq:PWW}), for small $\epsilon_{ij}$,
is given by
\begin{equation}
P(\nu_i\to \nu_j) \simeq |\epsilon_{ij}|^2\quad (i\neq j),\qquad
P(\nu_i\to \nu_i) \simeq 1-2\epsilon_{ii}
\end{equation}
The present bounds on the flavour violating $\epsilon_{ij}=\epsilon^*_{ji}$, coming 
from neutrino experiments, are
\begin{equation}\label{eq:nubounds}
|\epsilon_{\mu\tau}| < 0.013,\qquad
|\epsilon_{e\tau}| < 0.09,\qquad
|\epsilon_{e\mu}| < 0.05.
\end{equation}
The dominant bounds on $\epsilon_{\mu\tau}$ and $\epsilon_{e\tau}$ are due to the 
{\sc Nomad} experiment\footnote{The CERN neutrino
experiments, {\sc Nomad} and {\sc Chorus}, were motivated by theoretical prejudices for
small mixing angles and for warm dark matter.
Today they are among the most significant probes of extra-dimensional 
neutrinos.}~\cite{CHORUSeNOMAD},
$P(\nu_{\mu}\rightarrow\nu_{\tau})<1.6\times 10^{-4}$ and
$P(\nu_{e}\rightarrow\nu_{\tau})<7.4\times 10^{-3}$ at $90\%$ CL. 
The bound on $\epsilon_{e\mu}$ comes from {\sc Karmen}~\cite{Karmen} and {\sc Nomad}~\cite{CHORUSeNOMAD}. 
The LSND anomaly~\cite{LSND}, which may be interpreted as evidence for
$P(\bar{\nu}_\mu\to\bar{\nu}_e)\approx 2.5\times 10^{-3}$
could tentatively be solved if $|\epsilon_{e\mu}| \approx 0.05$.
This is in slight conflict with the bound from {\sc Karmen}, 
and in strong contradiction with the bounds from 
$\mu\rightarrow e\gamma$, which we will discuss in the next section.

The diagonal elements $\epsilon_{ii}=\epsilon^*_{ii}$ are constrained by {\sc Chooz}~\cite{CHOOZ} and Bugey~\cite{Bugey}, 
yielding $\epsilon_{ee} < 0.025$ and, because of the modified neutrino
couplings to the $Z$ boson, by the invisible $Z$ width measured at 
LEP~\cite{lepwg},
which leads to $|\epsilon_{ee} +\epsilon_{\mu\mu} +\epsilon_{\tau\tau}|
<0.013$ at 90{\%}CL\footnote{Since the present measurement of neutrino
counting at LEP agrees with the SM only at the 2-$\sigma$ level, and 
a nonzero $\epsilon$ reduces the effective number of neutrinos to 
$N_\nu =3-2\Tr \epsilon_{ij}$, this measurement gives a possible indication of 
a positive effect, $\epsilon_{ee} +\epsilon_{\mu\mu} +\epsilon_{\tau\tau}  
=0.008\pm 0.004$.}.
The $\ell_i W \nu$ couplings are a factor $1- \epsilon_{ii}/2$ smaller than in the SM, 
so that lepton universality tests~\cite{Pich} 
in $\tau$ and $\pi$  decays give
%%% can be  translated into the following $90\%$CL bounds
\begin{equation}\label{eq:CC}
\epsilon_{ee}-\epsilon_{\mu\mu}=0.0027\pm0.0024,\qquad
\epsilon_{\mu\mu}-\epsilon_{\tau\tau}=0.0003\pm0.0042,\qquad
\epsilon_{ee}-\epsilon_{\tau\tau}=0.0011\pm0.0045.
% |\epsilon_{\mu\mu}-\epsilon_{ee}| ,~
% |\epsilon_{\tau\tau}-\epsilon_{\mu\mu}|,~|\epsilon_{\tau\tau}-\epsilon_{ee}| < 0.007.
\end{equation}
A slightly more stringent bound, $\epsilon_{ee}+\epsilon_{\mu\mu} <0.002$, can be obtained by
comparing a global fit to the precision LEP data with the $\mu$ lifetime.
The bounds on the flavour violating $\epsilon_{ij}$ from
$Z,\tau,\mu,\pi$ decays are not competitive with the ones from neutrino oscillation
searches.

In the flat minimal model, the $\epsilon_{ij}$ parameters are determined 
in terms of the single 
unknown parameter $\epsilon$, defined in Eq.~(\ref{eq:def_epsilon}), and of the 
measurable (and already ``partially'' measured)
neutrino oscillation parameters as
\begin{equation}\label{eq:minimal}\begin{array}{ll}
\displaystyle
\epsilon_{\tau\tau}\approx \epsilon_{\mu\mu}\approx \epsilon_{\mu\tau} \approx\frac{\epsilon}{2},\qquad&
\displaystyle
\epsilon_{e e} \approx \left( |U_{e3}|^2 + \frac{\Delta m^2_{\sun}}{2\Delta m^2_{\atm}} \right)\epsilon , \\[5mm]
\displaystyle
\epsilon_{e \mu} \approx \left(\frac{|U_{e3}|}{\sqrt{2}} + 
\frac{e^{-i\phi}\Delta m^2_{\sun}}{2\sqrt{2}\Delta m^2_{\atm}} \right)\epsilon ,\qquad&
\displaystyle
\epsilon_{e \tau} \approx \left(\frac{|U_{e3}|}{\sqrt{2}} - 
\frac{e^{-i\phi}\Delta m^2_{\sun}}{2\sqrt{2}\Delta m^2_{\atm}} \right) 
\epsilon,
\end{array}
\end{equation}
where we assumed maximal solar and atmospheric mixing. 
Here $U_{e3}$ is the
element of the neutrino mixing matrix which is currently constrained 
to be small
by the {\sc Chooz} reactor neutrino data~\cite{CHOOZ}
and $\phi$ is the CP-violating phase.
The bounds previously discussed can be turned into constraints on $\epsilon$,
in the case of the minimal model. It turns out that
the most stringent bounds  are 
$\epsilon\circa{<}0.004$ (from $\mu,\tau,\pi$ decays and precision LEP data) and
$\epsilon\circa{<} 0.03$ (from neutrino experiments).

The most sensitive future experimental search
seems to be $\tau$ or wrong signed $\mu$ appearance at a future 
near-detector of a
$\nu$-factory. Such a  detector, located at $\circa{<} 1$ km from the 
neutrino source, has already
been studied in detail as a tool for neutrino deep inelastic scattering 
experiments~\cite{report}.
It expects to observe $10^8$ charged current SM events per year.
It seems possible~\cite{report} to probe 
$\epsilon_{\mu\tau}$ and $\epsilon_{e\tau} $ values down to the level of
$10^{-4}$, 
by looking for $\tau$ appearance, and $\epsilon_{e\mu}$ down to similar values
by looking for wrong sign muons, therefore
improving the present bounds by two orders of magnitude.
More dedicated studies, however, are still required. 
The sensitivity on the flavour diagonal $\epsilon_{\mu\mu}$ from searches
in the disappearence channel seem to be slightly worse, being limited by
the uncertainty on the incoming neutrino flux. 
The sensitivity on $\epsilon_{ee}$ should be weaker than the one to
$\epsilon_{\mu\mu}$, while $\epsilon_{\tau\tau}$ effects cannot
be probed.

\subsubsection*{CP violating effects}
In order to understand qualitatively what the
general exact formula Eq.\eq{PWW}  means in the various possible models,
it is useful to specialize it to the short-baseline case $\delta_{ij}\ll1$ where
$\delta_{ij}\equiv \Delta m^2_{ij} L/2E_\nu$.
Furthermore, it is useful to count how many CP-violating phases are contained in the
$\epsilon_{ij}=\epsilon_{ji}^*$ parameters.
\begin{itemize}
\item[1.] In processes where neutrino masses can be neglected,
the $\epsilon_{ij}$ contain $(N_g-1)(N_g-2)/2$ CP-violating phases (where $N_g=3$ is 
the number of generations) which do not give rise to any observable effects.

\item[2.] If the $\epsilon_{ij}$ can be neglected (normal oscillations),
the neutrino mixing matrix contains the usual $(N_g-1)(N_g-2)/2$ CP phases.
For $N_g=3$ there is one CP-violating phase, $\phi$,
and the CP-asymmetric part of the 
oscillation probability is the same in all flavour transitions $\nu_i\to \nu_j\neq \nu_i$,
and, for small $L$, is proportional to  $L^3$:
\begin{equation}
P_{\CPV}\equiv \frac{P(\bar\nu_i\to \bar\nu_j)-P(\nu_i\to \nu_j)}{2}=
 8J_{\rm CP} \sin\frac{\delta_{12}}{2} 
\sin\frac{\delta_{13}}{2}\sin\frac{\delta_{23}}{2}\approx 
J_{\rm CP}\delta_{12}\delta_{13} \delta_{23},
\end{equation}
where $J_{\rm CP}\approx -\frac{1}{2}U_{e3}\sin\phi$
(assuming maximal mixing in atmospheric and solar oscillations, and $|U_{e3}|\ll1$).

\item[3.]
In a generic process affected by neutrino masses and by the $\epsilon_{ij}$,
the neutrino mixing matrix contains the usual $(N_g-1)(N_g-2)/2$ CP phases,
and the $\epsilon_{ij}$ parameters contain other $N_g(N_g-1)/2$ CP phases.
Therefore, unlike the standard case, CP violation can be present
when only the dominant two-generation $\nu_\mu \leftrightarrow \nu_\tau$ 
atmospheric oscillations is ``turned-on.''
\end{itemize}
%In order to avoid long expressions, we will assume that $L$ small enough, so that 
%$\delta_{ij}\ll 1$. 
For example, ignoring subleading $\delta_{\rm sun}$ effects and keeping terms
up to second order in $\delta_{\rm atm}$
\begin{equation}
P(\nu_\mu\to \nu_\tau) = \left|\epsilon_{\mu\tau} - 
\sin 2 \theta_{\atm} \frac{i\delta_{\atm}}{2} \right|^2,\qquad
\frac{\delta_{\atm}}{2} = 10^{-4} \frac{\Delta m^2_{\atm}}{3\cdot
10^{-3}\eV^2}\frac{L}{\hbox{km}}\frac{20\GeV}{E_\nu}.
\end{equation}
The fact that these two effects could be comparable
offers an opportunity to measure CP-violating effects present, if $\epsilon_{\mu\tau}$ 
is complex, as a difference in $P(\nu_\mu\to \nu_\tau)$ from $P(\bar{\nu}_\mu\to 
\bar{\nu}_\tau)$ proportional to $L$. The CP asymmetry can be maximal with a suitable 
choice of the pathlength and neutrino energy, if $\epsilon_{\mu\tau}$ has a large 
CP-violating phase~\cite{GGGN}.

%We now want to discuss how large CP-violating effects can be in the specific models
%presented in the previous section.
In the minimal model presented in the previous section, the only CP-odd phase is the
one present in the neutrino mass matrix (recall that $\epsilon_{ij}\propto 
(m_{\nu}m_{\nu}^{\dagger})_{ij}$), and new CP-violating effects (proportional to $L$) 
are suppressed by  $J_{\rm CP}$  and by two powers of $\epsilon$:
\begin{equation}
P_{\CPV}=J_{\rm CP}[- 2 \delta_{12} \epsilon^2 + \delta_{12}\delta_{13} \delta_{23}].
\end{equation}
Given the current constraint on $\epsilon\circa{<} 0.01$, such effects are negligible.

\smallskip

For specific non minimal models, it is necessary to check whether the (indirect) relation
of $\epsilon_{ij}$ and the neutrino mass matrix allows to ``rotate away'' 
potentially physical phases in $\epsilon_{ij}$. 
This is the case if a single right-handed neutrino gives a dominant contribution to $\epsilon_{ij}$,
as happens in the minimal model.
Furthermore, in order to obtain
observable effects, it is necessary to generate a large $\epsilon_{ij}$ with a 
large phase, while keeping $|\epsilon_{e\mu}|$ small, due to the severe
constraints already imposed by searches for $\mu\to e\gamma$ (see next section).  
In spite of all the difficulties, however, it is possible to build specific
models which yield significant CP-violating effects. 

%In all the  models that we have considered
%CP violation disappears in the limit where
%only the physics responsible of the atmospheric mass splitting
%is kept.
%Therefore
%$P_{\CPV}$ is at most of order 
%``$\epsilon_{\sun} \delta_{\atm} - \epsilon_{\atm}\delta_{\sun}$''.
%The largest effects of order  ``$\epsilon_{\atm}\delta_{\atm}$'' allowed by the general
%low energy analysis at point 3.\ are not present.
%Nevertheless, $P_{\CPV}$ is not suppressed by $J_{\rm CP}\propto U_{e3}$, and
%no longer needs to be the same in all flavour transitions.
%Taking into account that the solar $\epsilon_{\sun}$ is severely constrained by 
%$\mu\to e \gamma$ (see next section),
%in these models the CP-asymmetry in $\nu_\mu\leftrightarrow \nu_\tau$ is at most a factor
%$\Delta m^2_{\rm sun}/\Delta m^2_{\rm atm}$ smaller than its maximal value discussed above.
%
%In the other $\nu_e\leftrightarrow \nu_\tau$ and $\nu_e \leftrightarrow \nu_\mu$ 
%transitions,
%the CP even part oscillation probability is suppressed
%by $U_{e3}$ or by $\Delta m^2_{\rm sun}/\Delta m^2_{\rm atm}$,
%but the CP odd part can be comparable to the one in $\nu_\mu\to \nu_\tau$.
%It is possible to build models that give the maximal CP-violating effects,
%e.g.\ using a third right-handed neutrino that contributes negligbly to neutrino masses
%and significantly to the tree-level operator Eq.\eq{TreeLevel}.

\subsubsection*{Matter effects}
Non-flavour diagonal interactions with the $Z,W$ bosons give rise to
non standard MSW corrections. The corresponding ``matter potential,'' however,
is too small to solve the solar neutrino puzzle or affect significantly the
SM effect due to charged current $\nu_e-e$ scattering.
On the other hand, the SM matter effects in $\nu_\mu \leftrightarrow\nu_\tau$ 
oscillations are suppressed by $\sim (m_\tau/\pi M_W)^2$~\cite{Marciano}, so that non SM effects 
could be dominant and perhaps detectable. In the minimal flat model,  
$\nu_\mu\leftrightarrow \nu_\tau$ oscillations in neutral matter are modified to 
\begin{equation}
P(\bar\nu_\mu\to \bar\nu_\mu) =1-\sin^2(2 \theta_{\atm})\sin^2 
\bigg[\bigg(\frac{\Delta m^2_{\atm}}{4E_\nu}- \frac{\epsilon}{8} 
\sqrt{2} G_FN_e\bigg)L\bigg]+
\Ord(\epsilon^2),
\end{equation}
and could result in a narrow resonance in long baseline oscillations of high energy 
anti-neutrinos, $E_\nu\sim 30\GeV/\epsilon$. 
Similar effects have been discussed in~\cite{gago}.

Matter effects are also known to substantially change the expected neutrino fluxes
from supernov\ae{}~\cite{Smirnov}.
The new matter effects due to the KK right-handed neutrinos 
produce a new MSW resonance in supernov\ae{} which seems, however, not to 
perturb the observable $\nu_e$ and $\bar{\nu}_e$ spectra in a significant way.

\setcounter{footnote}{0}
\setcounter{equation}{0}
\section{One-loop effects in charged lepton processes}

The effects of any high energy theory can be described at much lower energies
by  higher dimensional operators. The relevant set of operators and
their coefficients depend on the ultraviolet cut-off which is used to make computations in
the low energy effective theory. 
Using dimensional regularization (so that no power divergences arise in 
loop computations),
the low energy effects of bulk neutrinos are described by a
few dimension-six operators, including $\ell_i\ell_j \gamma$ magnetic moment operators, 
$\ell_i\ell_j Z$ vertices, and four fermion interactions.

These effects are partly due to computable SM loop corrections to the 
operator in Eq.\eq{TreeLevel}, generated at tree level. However, comparable contributions 
come from high energy effects in the full theory containing the KK modes of
the higher dimensional fermions. As before, it is important to emphasize that the 
coefficients of the effective operators cannot be computed, since we do not have a 
renormalizable high-energy theory. {\sl The coefficient of each operator is a 
free parameter} that cannot be calculated in terms of the non-renormalizable Yukawa 
couplings $\lambda$ of the extra-dimensional neutrino.
At best, they can be estimated by introducing some explicit cut-off $\Lambda$.
It proves particularly useful to rewrite the dimensionful Yukawa coupling ${\lambda}$ as
\begin{equation}
{\lambda} \equiv \frac{\bar\lambda}{\Lambda^{\delta/2}\sqrt{\ell_\delta}},
\end{equation}
because we can interpret the dimensionless 
couplings $\bar{\lambda}$ as parameters for understanding how ``strongly coupled''
the right-handed neutrinos are.\footnote{When in section~2 we discussed the 
connection of neutrino masses with gravity, 
it was convenient to parameterize
the dimensionful fundamental Yukawa coupling $\lambda$
in terms of another  dimensionless
coupling $\hat{\lambda}$, defined using the reduced Planck mass as the unit of mass.}
As defined in Eq.~(\ref{pitpit}), 
$\ell_\delta$ is a $\delta$-dimensional loop factor
(for example $\ell_4\sim 1/(4\pi)^2$).  All virtual effects
are comparable to the effects of a four-dimensional heavy right-handed neutrino
with a Yukawa coupling $\bar\lambda$
(so that $\bar\lambda\sim 4\pi$ corresponds to strong coupling).
The effective Lagrangian at energies smaller than the cut-off $\Lambda$ is 
\begin{equation}
{\cal L} \sim \frac{\bar\lambda^2}{\Lambda^2} 
(\bar{L}H^* \ds L H) + \frac{\ell_4}{\Lambda^2} \left\{
e \bar\lambda^2 
[\mu \to e \gamma] +  \bar\lambda^2(\bar\lambda^2+g^2) \left(
g v^2 [Z\mu e]+ 
[\mu\to 3 e]\right) + \bar\lambda^2 g^2 [\mu q\to e q] \right\} . 
\label{eq:effective_L}
\end{equation}
We have used a shorthand notation to indicate the various 
${\rm SU}(2)_L\otimes{\rm U}(1)_Y$-invariant operators. For example, 
$g v^2 [Z\mu e]$ indicates operators like $ (H^* D_\mu H)(\bar{L} \gamma_\mu L)$. 
While the magnetic operator $[\mu \to e \gamma]$ is generated at order
$\bar\lambda^2$, other dimension-six operators also receive contributions
at order $\bar\lambda^4$. In the case of the minimal model
presented in Sec.~2, the loop effects are explicitly calculated (making use of a hard
ultraviolet cut-off) in the Appendix.
 
% Note that all dimension-six operators decouple as 
% $1/\Lambda^2$ when $\Lambda\rightarrow\infty$. This is in contrast with the
% claim of non-decoupling effects made in~\cite{Pilaftsis}. The decoupling
% actually holds also in the expressions presented in~\cite{Pilaftsis}, but
% it is hidden inside the definition of the mixing parameters. 

Note that all dimension-six operators decouple as
$1/\Lambda^2$ when $\Lambda\rightarrow\infty$. The  correct decoupling
is also found in the expressions presented in ref.~\cite{Pilaftsis}, once
the definition of the mixing parameters is taken into account.

We remark that we are only considering the effects due to the
extra-dimensional right-handed neutrinos: a full quantum gravity theory at TeV energies 
is expected to give additional effects. In particular, it is
important to note that
while some unknown mechanism could be responsible for suppressing effects which
lead, {\it e.g.,}\/ to proton decay, it is hard 
to believe that there are no extra contributions to lepton flavour violating 
processes, given that the neutrino data indicates that individual lepton
flavours are strongly broken symmetries 
(see~\cite{models} for a discussion of these effects and 
possible ways of addressing this issue). Nonetheless, we will ignore here such 
contributions, which are impossible to estimate.

In spite of all the intrinsic uncertainties,  useful results can be obtained.
Both the tree level and the more relevant one-loop operators are 
suppressed by the same order
 of $\bar{\lambda}/\Lambda$. This implies that one-loop effects are only 
suppressed by a factor of order $1/16\pi^2$ with respect to the tree level 
term.\footnote{This nice property is not shared by virtual effects mediated by
higher dimensional gravitons,
that generate dimension 6 operators at loop level but 
only dimension 8 operators at tree level.}
Some operators, like the four-fermion operators contributing to
$\mu\rightarrow 3e$ and $\mu\to e$ conversion in nuclei, may in fact be
enhanced with respect to $\mu \to e \gamma$
by $\bar{\lambda}^2/g^2$ if $\bar{\lambda}> g$.
Moreover, the coefficients of the operators included in $[Z\mu e]$ 
and $[\mu \to 3e]$ receive a $\ln \Lambda^2 /m_W^2$ enhancement.  
Extra-dimensional models predict (up to order one factors)
 relations between flavour violating effects in the charged and neutral lepton sectors. 
In the simplest ``flat'' model we discussed in Sec.~2, 
predictions for many charged lepton flavour violating processes will be
strictly related to the observed neutrino oscillation parameters ($\Delta m^2$, and 
neutrino mixing angles), such that, {\it e.g.}, it is possible to predict 
${\rm BR}(\tau\rightarrow\mu\gamma)/{\rm BR}(\mu\rightarrow e\gamma)$.

Therefore, minimal extra-dimensional models are more predictive than other 
beyond-the-SM sources of lepton flavour violating phenomena related to right 
handed neutrinos.
In the constrained MSSM, for example, the presence of very heavy right-handed 
neutrinos (that generate small neutrino masses via the see-saw mechanism) 
yields potentially large flavour violating effects in the muon sector. While all 
branching ratios are precisely calculable in terms of the parameters of the theory 
(the MSSM is a perturbative gauge field theory), it is impossible to establish a 
connection between the observed flavour-violating neutrino
masses  with predictions for $\mu\rightarrow e\gamma$, etc, due to the presence of too 
many unknown flavour-mixing parameters in the right-handed neutrino sector.

\smallskip

In this section, we concentrate on charged lepton flavour violating phenomena, and
also comment on the anomalous magnetic moment of the muon. 
Note, however, that the effective Lagrangian Eq.~(\ref{eq:effective_L}) also allows 
$e^+e^-\rightarrow \ell_i^+\ell_j^-$
($i,j=e,\mu,\tau$) processes~\cite{Pilaftsis}.
Such effects could be studied at the $Z$ resonance with a next-generation 
$e^+e^{-}$ collider. However, searches for
processes with $i\neq j$ do not seem very promising given the current 
bounds on $\epsilon_{ij}$~\cite{Vissani},
and the effects on processes with $i=j$ seem less sensitive than
corrections to $Z$ observables, which are generated at tree level.

Explicit expressions for the processes of interest are derived and listed in 
the Appendix. They agree, apart from  minor differences, with the 
corresponding expressions in~\cite{Pilaftsis} when the 
same model and the same 
arbitrary cut-off is chosen. Our numerical results are, instead, different. 
In particular, the experimental bounds we obtain are much weaker
and do not, in general, require 
fundamental scales above $100\TeV$ or $10\TeV$. 
We discuss some of our results in what follows.

\subsection*{Flat extra dimensions}

We will concentrate on the minimal model outlined in Sec.~2. 
This model contains only two new free parameters, $\epsilon$ and $\Lambda$.
Up to order one factors, the rates for the different charged lepton process depend only 
on these two parameters, on the neutrino masses, and on the elements of the
standard neutrino mixing matrix $U_{\alpha i}$. 
The largest Yukawa coupling is then determined by the relation
\begin{equation}
\bar \lambda =\frac{\Lambda}{v}\sqrt{\epsilon (\delta -2)} .
\end{equation}
As mentioned before, we will
assume that the neutrino masses are hierarchical $(m_1^2\ll m_2^2\ll m_3^2)$ such that
all information regarding neutrino parameters can be obtained from neutrino oscillation
experiments. 
We later comment on non minimal flat models and warped models.

\subsubsection*{$\mu\rightarrow e\gamma$ and $\tau\rightarrow\mu\gamma$}

Using the expressions derived in the Appendix, the branching ratio for  
$\mu\rightarrow e\gamma$ can be written as follows:
\begin{eqnarray}
{\rm BR}(\mu \rightarrow e \gamma) &=&
\frac{3 \alpha}{8\pi}
%\left[ \frac{S_\delta}{\delta-2} 
%\left(\frac{\Lambda}{M_F}\right)^\delta
%\right]^2 
\epsilon^2
\left|
\sum_j U_{ej} U_{\mu j}^* \frac{m^2_{j}}{\Delta m^2_{\atm}}
\right|^2 =  \frac{3 \alpha}{8\pi} \epsilon^2
\left|U_{e2} U_{\mu 2}^* \frac{\Delta m^2_{\sun}}{\Delta m^2_{\atm}}
+U_{e3} U_{\mu 3}^* 
\right|^2.
\label{meg}
\end{eqnarray}
Similarly, the branching ratio for $\tau \rightarrow \mu \gamma$ is
given by
\begin{eqnarray}
{\rm BR}(\tau \rightarrow \mu \gamma)
&=& 
0.174 \frac{3 \alpha}{8\pi}
\epsilon^2
\left|
\sum_j U_{\mu j} U_{\tau j}^* \frac{m^2_{j}}{\Delta m^2_{\atm}}
\right|^2
= 0.174 \frac{3 \alpha}{8\pi} \epsilon^2
\left|U_{\mu 2} U_{\tau 2}^* \frac{\Delta m^2_{\sun}}{\Delta m^2_{\atm}}
+U_{\mu 3} U_{\tau 3}^* 
\right|^2.
\label{tmg}
\end{eqnarray}
The branching ratio for $\mu \rightarrow e \gamma$ depends on the solution to the
solar neutrino puzzle (i.e. $U_{e2}$ and $\Delta m^2_{\sun}$)
and on the unknown $U_{e3}$ element. On the other hand, 
${\rm BR}(\tau \rightarrow
\mu \gamma)$ depends dominantly on the well measured atmospheric 
parameters:\footnote{
The $\tau \rightarrow e \gamma$ branching ratio is suppressed by the unknown
reactor angle $U_{e3}$, so that it is smaller and more uncertain than 
the $\tau \rightarrow \mu \gamma$ branching ratio.} 
$|U_{\mu 3} U_{\tau 3}| \approx 0.5$ and $\Delta m^2_{\atm}\approx 
3\times 10^{-3}\eV^2$, such that
\begin{eqnarray}
{\rm BR}(\tau \rightarrow \mu \gamma) \simeq 0.174
\frac{3 \alpha}{8 \pi} \epsilon^2 
\left|U_{\mu 3} U_{\tau 3}^*\right|^2 =
4 \times 10^{-5} ~\epsilon^2.
\label{tmg_model2}
\end{eqnarray}
The bound Eq.\eq{nubounds} from neutrino experiments on $\epsilon_{\mu\tau}\approx 
\epsilon/2$ implies, in a model-independent way, that a possible effect in  
$\tau \rightarrow \mu \gamma$ is at least two  orders of magnitude below 
the present limit ${\rm BR}(\tau \rightarrow \mu \gamma)< 1.1 \times 10^{-6}$.

The dependence on the free parameter $\epsilon$ cancels out in the
ratio between ${\rm BR}(\mu \rightarrow e \gamma)$ and 
${\rm BR}(\tau \rightarrow \mu \gamma)$, that
is expressed only in terms of neutrino oscillation parameters,
\begin{eqnarray}
\frac{{\rm BR}(\mu \rightarrow e \gamma)}
{{\rm BR}(\tau \rightarrow \mu \gamma)}
&=& 
 \frac{\left|
U_{e2} U_{\mu 2}^* \Delta m^2_{\sun}
+U_{e3} U_{\mu 3}^* \Delta m^2_{\atm} \right|^2}
{0.174 \left| U_{\mu 3} U_{\tau 3} \Delta m^2_{\atm} \right|^2}.
\end{eqnarray}
The numerical value depends on the still unmeasured $U_{e3}$ and on which is the true 
solution of the solar neutrino puzzle.
The dominant contribution to $\mu\to e \gamma$ could be related to the ``atmospheric'' or 
``solar'' neutrino mass splitting. If the solar anomaly is due to LMA oscillations, 
and if $|U_{e3}|$ is such that 
$\Delta m^2_{\sun} \gg U_{e3} \Delta m^2_{\atm}$, the ``solar'' contribution 
dominates, and we predict
\begin{equation}
{\rm BR}(\mu \rightarrow e \gamma) \approx 1\times 10^{-8}
\epsilon^2 \left( \frac{\Delta m^2_{\sun}}{3\times10^{-5}~{\rm eV}^2}\right)^2,
\end{equation}
\begin{equation}
\frac{{\rm BR}(\mu \rightarrow e \gamma)}
{{\rm BR}(\tau \rightarrow \mu \gamma)} \approx
3 \times 10^{-4} \left(\frac{|U_{e2}U_{\mu 2}|}{0.35}\frac{0.5}{|U_{\mu 3}U_{\tau 3}|}
\right)^2  
\left( \frac{\Delta m^2_{\sun}}{3\times10^{-5}~{\rm eV}^2} 
\frac{3\times 10^{-3}~{\rm eV}^2}{\Delta m^2_{\atm}}
\right)^2.
\end{equation}
When compared to the current experimental bound 
${\rm BR}(\mu \rightarrow e \gamma)<1.2\times 10^{-11}$, we obtain the limit
\begin{equation}
\epsilon < 3\times 10^{-2}\left( \frac{3\times10^{-5}~{\rm eV}^2}
{\Delta m^2_{\sun}}\right) ~~~~{\rm if}~~|U_{e3}|\ll 
{\Delta m^2_{\sun} \over \Delta m^2_{\atm}}.
\label{lim1} 
\end{equation}
In the near future the experimental sensitivity to BR$(\mu\to e\gamma)$
is expected to reach $10^{-14}$~\cite{barkov}. 

On the other hand, if the solar parameters fall in the LOW or SMA 
regions,\footnote{After SNO~\cite{SNO}, the SMA solution is strongly disfavoured by 
the solar neutrino data~\cite{postSNO}.} 
and $|U_{e3}|$ is large enough, the ``atmospheric'' contribution 
to ${\rm BR}(\mu \rightarrow e \gamma)$ dominates, so that
\begin{equation}
{\rm BR}(\mu \rightarrow e \gamma) \approx 4\times 10^{-5}~\epsilon^2 \, |U_{e3}/0.3|^2,\qquad
\frac{{\rm BR}(\mu \rightarrow e \gamma)}
{{\rm BR}(\tau \rightarrow \mu \gamma)}
\approx   \left|\frac{U_{e3}/0.3}{U_{\tau 3}/0.7} \right|^2.
\end{equation}
In this case the bound on $\epsilon$ becomes
\begin{equation}
\epsilon < 5\times 10^{-4} |0.3/U_{e3}| ~~~~{\rm if}~~|U_{e3}|\gg 
\Delta m^2_{\sun}/\Delta m^2_{\atm}.
\label{lim2} 
\end{equation}
If  $U_{e3}$ is close to its current experimental upper bound, $|U_{e3}|\circa{<}0.3$,
the bound on $\epsilon$ is so strong that 
experimental signals in the neutrino sector can only be very small.

Fig.~\ref{fig:br} shows the values of ${\rm BR}(\mu\rightarrow e\gamma)$ and
${\rm BR}(\tau\rightarrow \mu\gamma)$ as a function of $|U_{e3}|$ for 
$\Delta m^2_{\sun}/\Delta m^2_{\atm}=10^{-2}$, $|U_{\mu3}/U_{\tau3}|^2=1$,
$|U_{e2}/U_{e1}|^2=2/3$
(as suggested by atmospheric and LMA solar oscillations),
 and no CP violation in the neutrino mixing matrix,
for different values of $\epsilon$ (and $\Lambda$). 
If the ``solar'' and ``atmospheric''
contributions to $\mu\rightarrow e\gamma$ are comparable (which happens at 
$U_{e3}\sim 10^{-(3\div 2)}$ for LMA solar oscillations), the
CP-violating phase in the neutrino mixing matrix may either enhance or suppress the 
branching ratio, depending on how the two terms interfere.
\begin{figure}
$$ \includegraphics[width=16cm]{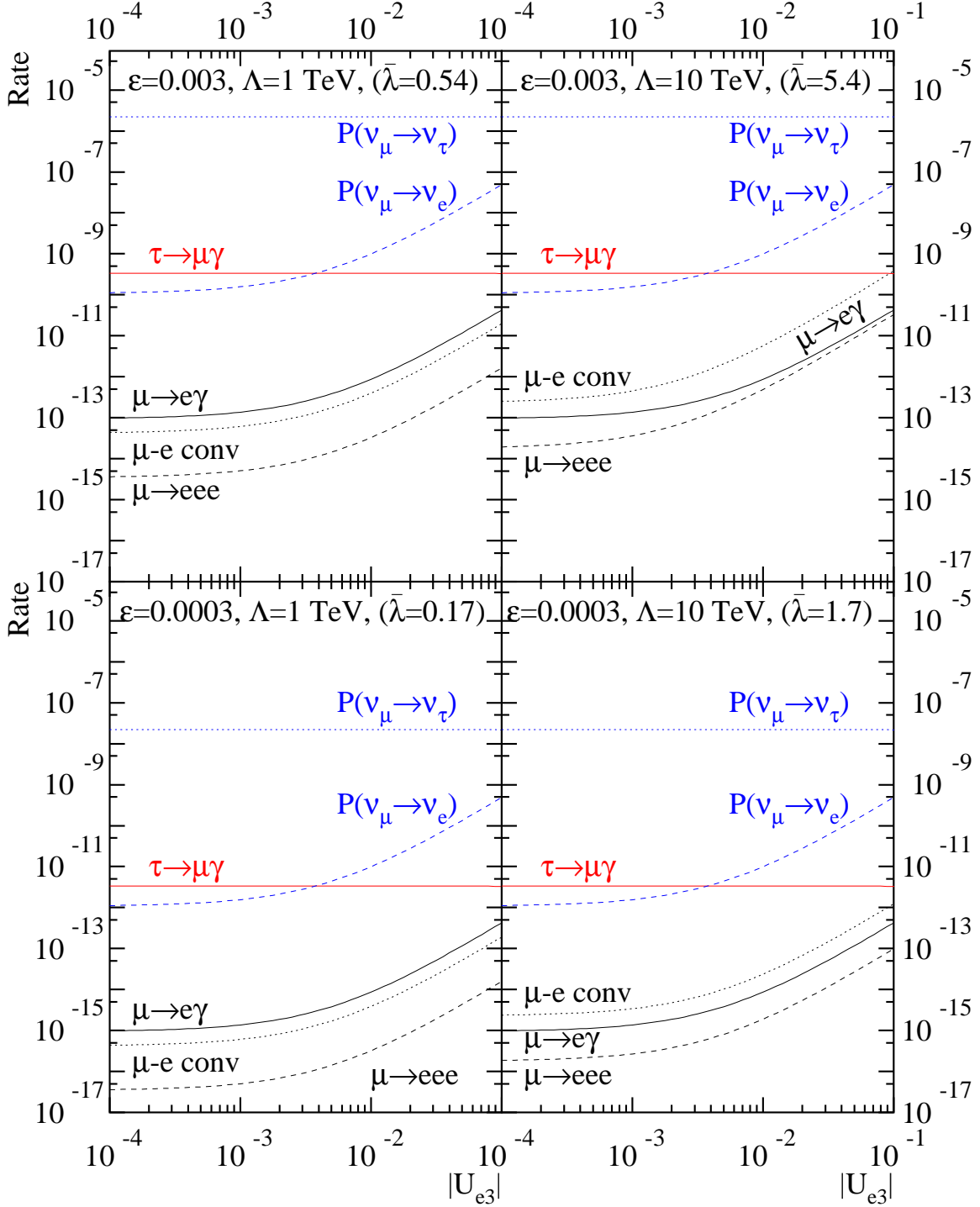} 
$$
  \caption{\em The most interesting leptonic observables as a function of $|U_{e3}|$, for $\delta=5$,
$\epsilon=0.003$ (top) or 0.0003 (bottom), and $\Lambda=1$~TeV (left) or 10~TeV (right). 
We assume $\Delta m^2_{\sun}/\Delta m^2_{\atm}=10^{-2}$, $|U_{\mu3}/U_{\tau3}|^2=1$,
$|U_{e2}/U_{e1}|^2=2/3$ ({\it i.e.,}\/ 
maximal mixing in the atmospheric sector and the LMA
solution to the solar neutrino puzzle), no
CP-violation in the neutrino mixing matrix, and
hierarchical neutrino masses ($m_1^2\ll m_2^2\ll m_3^2$). 
Note that $\bar{\lambda}$ is the largest Yukawa coupling,
$P(\nu_i\rightarrow\nu_j)$ is the neutrino 
conversion probability at very short baselines, and the $\mu\to e$ conversion
rate is computed for $^{27} {\rm Al}$.}
  \label{fig:br}
\end{figure}

%%%%%%%%%%%%%%%%%

\subsubsection*{$\mu\rightarrow e\bar{e}e$ and $\mu\rightarrow e$ conversion in nuclei}

Unlike the $\ell_i\rightarrow \ell_j\gamma$ decays, 
these rare leptonic processes are more dependent on the unknown ultraviolet details 
of the models, but still a few interesting results can be obtained.

In $\mu \rightarrow e\bar{e}e$ process, several operators may contribute significantly
(see the Appendix for detailed expressions).
First of all, the $\mu \rightarrow e \gamma$ magnetic penguin operator contribution to 
$\mu \rightarrow e\bar{e}e$ is enhanced by
$\ln(m_\mu/m_e)$, which is a consequence of a collinear divergence of the 
electron-positron pair in the $m_e\rightarrow 0$ limit.
Second, the $Z$ and photon penguin diagrams are enhanced by an ultraviolet divergence
$\sim \ln(\Lambda/m_W)$. When $\Lambda$ is significantly larger than 
$m_W$, this log-enhancement is important. Finally, as discussed after 
Eq.~(\ref{eq:effective_L}), the $Z$-penguin and
box contributions contain $\bar{\lambda}^4$-terms. 
If the higher-dimensional KK neutrinos are coupled  strongly enough 
such that $\bar{\lambda}>g$, then the
$\bar{\lambda}^4$-terms may dominate over the 
other contributions to BR$(\mu \rightarrow e\bar{e}e)$.

In light of this discussion, the ratio of branching ratios 
BR$(\mu \rightarrow e\bar{e}e)$/BR$(\mu \rightarrow e \gamma)$, in 
which the overall $\epsilon^2$ dependence in both processes cancels out,
can be written as 
\begin{equation}
\frac{{\rm BR}(\mu \rightarrow e\bar{e}e)}{ {\rm BR}(\mu \rightarrow e \gamma)}
\approx C_{\mu \rightarrow e \gamma}
+C_{\ln \Lambda} + C_{\bar{\lambda}}.
\label{m3e_to_meg}
\end{equation}
Here, $C_{\mu \rightarrow e \gamma}$ is the contribution from the 
$\mu \rightarrow e \gamma$ magnetic penguin operator
\begin{equation}
C_{\mu \rightarrow e \gamma}=\frac{\alpha}{3\pi}\left(
\ln \frac{m^2_\mu}{m^2_e}-\frac{11}{4} \right) = 6\times 10^{-3}.
\end{equation}
The second term $C_{\ln \Lambda}$ contains the $\ln \Lambda/m_W$ 
contributions from the $Z$ and photon penguin diagrams of $\mu \rightarrow e\bar{e}e$. 
For example, the $Z$-penguin diagram leads to
\begin{equation}
C_{\ln \Lambda} \sim \frac{3\alpha}{8\pi}
\left(3-\frac{2}{\sin^2\theta_W}+\frac{1}{2 \sin^2 \theta_W}\right)
\left( \ln\frac{\Lambda^2}{m^2_W}-\frac{2}{\delta-2}-\frac{5}{3}\right)^2
\sim 10^{-3}\left( \ln\frac{\Lambda^2}{m^2_W}-\frac{2}{\delta-2}-
\frac{5}{3}\right)^2 .
\end{equation}
We find that $C_{\ln \Lambda}=0.03$ ($0.2$)
for $\delta=5$ and $\Lambda=1$ TeV ($10$ TeV), when all 
ln terms present in the $Z$ and photon penguin diagrams are included.

The last term in Eq.(\ref{m3e_to_meg}), $C_{\bar{\lambda}}$, contains the contributions
from $\bar{\lambda}^4$ terms in the $Z$-penguin and box diagrams:
\begin{equation}
C_{\bar{\lambda}}\propto \left(\frac{\bar{\lambda}}{g}\right)^4 
\frac{\left| U_{e2}U_{\mu2}^* 
({\Delta m^2_{\sun}}/{\Delta m^2_{\atm}})^2+U_{e3}U_{\mu3}^*\right|^2}
{\left|U_{e2}U_{\mu2}^* 
({\Delta m^2_{\sun}}/{\Delta m^2_{\atm}})^{\phantom{2}}+U_{e3}U_{\mu3}^*\right|^2}.
\end{equation}
When $|U_{e3}|$ is small, $C_{\bar{\lambda}}$ is small (suppressed by the small
ratio of neutrino mass-squared differences squared).  On the other hand, if
$|U_{e3}|$ is sufficiently large, $C_{\bar{\lambda}}$ may in fact be the dominant
contribution to $\mu\rightarrow e\bar{e}e$ due to the potentially large $(\bar{\lambda}/g)^4$
enhancement.
Numerically, $C_{\bar{\lambda}}=0.6$ for $\Lambda=10$~TeV, $|U_{e3}|=0.1$,
and $\epsilon=0.003$, which corresponds to $\bar{\lambda}=5.4$.

Because of the contributions
$C_{\ln \Lambda}$ and $C_{\bar{\lambda}}$, the ratio
BR$(\mu \rightarrow e\bar{e}e)$/BR$(\mu \rightarrow e \gamma)$ can be
significantly larger than $10^{-1}$ if $\Lambda\circa{>}1$~TeV or $\bar{\lambda}>g$.
This is very different from predictions for lepton flavour violating processes
from SUSY models with slepton flavour mixing, in which the contribution
$C_{\mu \rightarrow e \gamma}$ is almost always dominant~\cite{KO_rev} and therefore
BR$(\mu \rightarrow e\bar{e}e)$/BR$(\mu \rightarrow e \gamma)\simeq 6\times 10^{-3}$. 
Large contributions to BR$(\mu \rightarrow e\bar{e}e)$
may be obtained in SUSY models with $R$-parity violation~\cite{SUSY_RP}.

The $\mu \rightarrow e$ conversion rate in nuclei behaves
similarly to the $\mu \rightarrow e\bar{e}e$ branching ratio: for
$\Lambda \gg m_W$, the log-enhancement in $Z$ and photon penguin diagrams is 
important and perhaps dominant, while if $\bar \lambda >g$, the 
$\bar{\lambda}^4$ term in the $Z$-penguin and box contributions
can be significant.
In both $\mu \rightarrow e\bar{e}e$ and $\mu \rightarrow e$ conversion in nuclei,
the $Z$-penguin contribution tends to dominate  because
of the $\ln \Lambda$ and
$\bar{\lambda}$ terms. When this is the case, the 
ratio R$(\mu \rightarrow e)$/BR$(\mu \rightarrow e\bar{e}e)$ 
does not depend on the unknown ultraviolet
details of the models. We verified numerically that
the ratio is almost constant, varying in the range 10--13 in the case of $\mu\to e$
conversion in $^{27}$Al, and 20--25 in $^{48}$Ti, 
in a large region of the parameter space. 
This feature 
%, which can also be observed in Fig.~\ref{fig:br},
is a definite prediction of the models under consideration.
Again, the situation is different from SUSY models with slepton flavor mixing, where
one expects R$(\mu \rightarrow e)$/BR$(\mu \rightarrow e\bar{e}e)\circa{<}1$.
%These ratios between event rates will be
%important quantities to distinguish different models of lepton
%flavour violation. 

Figure~\ref{fig:br} shows the branching ratio of $\mu\rightarrow e\bar{e}e$ and the
rate for $\mu\to e$ conversion in $^{27}$Al as a function of $|U_{e3}|$
for different values of $\Lambda$ and $\epsilon$ (we also fix $\delta=5$). 
All of the features we have discussed can be readily observed. First, one can 
clearly see that R$(\mu \rightarrow e ~{\rm in~^{27}Al})$/BR$(\mu 
\rightarrow e\bar{e}e)$ is roughly constant
($\sim 10$) for all the depicted values of $\epsilon$ and $\Lambda$. Second, at small
values of $|U_{e3}|$ (where the $\bar{\lambda}^4$ terms are negligible), the effect
of the $\ln(\Lambda/m_W)$ enhancement is visible: at $\Lambda=1$~TeV, 
BR$(\mu\to e\gamma)>$R$(\mu \rightarrow e~{\rm in~^{27}Al})$, 
while at $\Lambda=10$~TeV the situation 
is reversed. Finally, the $\bar{\lambda}^4$ enhancement is also clear if one compares  
R$(\mu \rightarrow e~{\rm in~^{27}Al})$/BR$(\mu\to e\gamma)$ and 
BR$(\mu \rightarrow e\bar{e}e)$/BR$(\mu\to e\gamma)$ at small and large values
of $|U_{e3}|$, for $\bar{\lambda}$ large. This behavior is more easily observed
for $\epsilon=0.003$ and $\Lambda=10$~TeV. Indeed, we have verified that
at even larger values of 
$\bar{\lambda}$, BR$(\mu \rightarrow e\bar{e}e)$ exceeds BR$(\mu\to e\gamma)$.
Note also that, for all values of $\epsilon$ and $\Lambda$ depicted in 
Fig.~\ref{fig:br}, the rate for $\mu\to e$ conversion in $^{27}$Al is larger than 
the proposed sensitivity reach of the MECO experiment~\cite{MECO}, even in the limit 
of small $|U_{e3}|$ (assuming the LMA solution to the solar neutrino puzzle).

Finally, we point out that, in general, the ``atmospheric'' and ``solar'' contributions
to $\mu\to e\bar{e}e$ have different CP-violating phases. Furthermore, 
their relative weight in the magnetic penguin operator 
is different from their relative weight in the four-fermion operators in 
Eq.\eq{effective_L}, unless $\bar{\lambda}\ll g$.
If the ``atmospheric'' and ``solar'' contributions are comparable,
the interference between them produces observable
CP-violating effects in polarized $\mu\to e\bar{e}e$ decays~\cite{CP in mu->3e}.

\subsubsection*{$g-2$ of the muon}
As derived in the Appendix (see also~\cite{Ng}), the contribution of the KK tower of
neutrinos to the muon anomalous magnetic moment is 
\begin{eqnarray}\label{eq:g-2}
\delta a_\mu &=& -\frac{g^2}{32 \pi^2} \frac{m^2_\mu}{m^2_W}\epsilon
\sum_{j} \left|U_{\mu j} \right|^2 \frac{m^2_{\nu j}}{\Delta m^2_{\atm}} 
\simeq - 10^{-9}\epsilon .
\end{eqnarray}
Here we have assumed maximal mixing in the solar sector and neglected $U_{e3}$ and 
$\Delta m^2_{\sun}/\Delta m^2_{\rm atm}$ corrections.
Note that the sign of the effect is {\sl negative,}\/ in contrast to the
BNL experimental result~\cite{g-2}, that claims a discrepancy $\delta a_{\mu}=+(4.3\pm1.6)\times 10^{-9}$
with respect to SM predictions (afflicted, however, by significant
hadronic uncertainties).
% The bound on $\epsilon$ is not particularly strong, however. 
%  At the three sigma level,
%  $\delta a_{\mu}\circa{>}-0.5\times 10^{-9}$, which implies $\epsilon\circa{<}0.5$. 
On the other hand, the present bounds on the $\epsilon_{ij}$ parameters
already require in a model-independent way that
the right-handed neutrino correction to $g-2$
is smaller than the theoretical uncertainty in the hadronic SM contributions.
%  
%  It should be noted, however, that the contribution of KK neutrinos is smaller
%  than future error estimates for $\epsilon\circa{<}0.1$, which is already 
%  required by other experimental bounds. 
If, in the future,  more precise experimental and theoretical results
establish the presence of a non SM correction to $g-2$, extra dimensional
models could still account for it by invoking ad-hoc dimension-six operators like
$\frac{1}{\Lambda^2}(\bar{L} \tau^a\sigma_{\mu\nu} D\hspace{-1.5ex}/~ L)W^a_{\mu\nu}$
with
$\Lambda \approx (3\div 4)\,{\rm TeV}$.
There is no direct contradiction between such 
non renormalizable  operators and 
bounds from the more sensitive precision data.  %~\cite{LEPparadox}.

\subsection*{Non minimal flat models and warped models} 
Nonminimal flat models and models with warped extra dimensions do not allow one
to relate the coefficients of the dimension-six effective operators to the parameters
in the neutrino Dirac mass matrix. For this reason, it is not possible to make 
interesting predictions. 
In non minimal flat models (B) and in warped models the na\"{\i}ve expectation 
is that all 
$\epsilon_{ij}$ are comparable,
but is easy to avoid this conclusion.
It is useful to estimate the $\ell_i\to \ell_j \gamma$ decays in terms of the
$\epsilon_{ij}$ parameters defined in Eq.~(\ref{eq:TreeLevel}). 
Up to order one uncomputable factors, $\ell_i\to \ell_j \gamma$ experiments give the 
following bounds 
\begin{equation}\label{eq:leptonbounds}
|\epsilon_{e \mu} |\circa{<} 10^{-4},\qquad
|\epsilon_{e\tau}|,|\epsilon_{\mu\tau}|\circa{<} 10^{-1}.
\end{equation}
Such bounds prohibit the interpretation of 
the LSND anomaly~\cite{LSND} as due to $\epsilon_{e\mu}$, 
as discussed in the previous section. 
In fact, the present bound on $\epsilon_{e\mu}$ is so strong that it will be hard 
to observe its effects in $\nu_e\leftrightarrow\nu_{\mu}$ transitions, even 
with a neutrino factory.

\begin{table}[t]
$$
\begin{array}{|c|lclc|lclc|}\hline
& 
\multicolumn{4}{|c|}{\hbox{\Blue present {\bf bounds} from experiments with\Black}} & 
\multicolumn{4}{|c|}{\hbox{\Blue future {\bf sensitivity} from experiments with\Black}} \\ \hline
&
\multicolumn{2}{|c}{\hbox{\Blue neutrinos\Black}}&
\multicolumn{2}{|c}{\hbox{\Blue charged leptons\Black}}&
\multicolumn{2}{|c}{\hbox{\Blue neutrinos\Black}}&
\multicolumn{2}{|c|}{\hbox{\Blue charged leptons\Black}}\\ \hline
\Blue|\epsilon_{e\mu}|\Black &  
<0.05 &  \hbox{({\sc Karmen})} &
\circa{<} 10^{-4} &  (\mu\to e \gamma) &
 \sim 10^{-4} &  \hbox{($\nu$ factory)} &
 \sim 10^{-5\div 6} &  (\mu~{\rm decays}) \\
\Blue|\epsilon_{e\tau}|\Black &  
<0.09 &  \hbox{(NOMAD)} &
\circa{<} 10^{-1} &  (\tau\to e \gamma) &
 \sim 10^{-4} &  \hbox{($\nu$ factory)} &
 \sim 10^{-1\div 2} &  (\tau\to e\gamma) \\
\Blue|\epsilon_{\mu\tau}|\Black &  
<0.013 &  \hbox{(NOMAD)} &
\circa{<} 10^{-1} &  (\tau\to \mu \gamma) &
 \sim 10^{-4} &  \hbox{($\nu$ factory)} &
 \sim 10^{-1\div 2} &  (\tau\to\mu\gamma) \\
\Blue\epsilon_{ee}\Black &  
<0.025 &  \hbox{(reactors)} &
\circa{<} 10^{-3} &  (Z\hbox{ data}) &
 \sim 10^{-3} &  \hbox{($\nu$ factory)} &
 \sim 10^{-3\div 4} &  (Z\hbox{ data}) \\
\Blue\epsilon_{\mu\mu}\Black &  
~~~\hbox{---} &&
\circa{<} 10^{-3}&  (Z\hbox{ data}) &
 \sim10^{-3} &  \hbox{($\nu$ factory)} &
 \sim 10^{-3\div4} &  (Z\hbox{ data}) \\
\Blue\epsilon_{\tau\tau}\Black &  
~~~\hbox{---}  &&
\circa{<} 10^{-2} &  (Z\to\nu\bar\nu) &
~~~\hbox{---} &&
 \sim 10^{-3} &  (Z\to\nu\bar\nu) \\ \hline
\end{array}$$
\caption[1]{\em Bounds  and future sensitivity reaches for the $\epsilon_{ij}$ coefficients defined by Eq.\eq{TreeLevel}.}\end{table}

% \parbox[c]{2cm}{\centering \Blue Goodness\\ of fit\Black} & 
% \hbox{\Blue ($a$) Rates only} &
% \parbox[c]{4cm}{($b$) Rates and spectra:\\\phantom{(b) }naive result}&
% \parbox[c]{4cm}{($c$) Rates and spectra:\\ \phantom{(c) }refined result}\Black \\ \hline

\setcounter{equation}{0}
\section{Conclusions}
We have shown that the various minimal and non minimal models with right-handed neutrinos 
in flat or warped extra dimension are described at  low energy by neutrino masses plus 
a specific set of dimension six operators. Up to order one factors (that are anyhow 
uncomputable since these models are not renormalizable),
flavour conserving effects are described by three positive numbers $\epsilon_{ee},~\epsilon_{\mu\mu},~\epsilon_{\tau\tau}$,
and flavour-violating effects are described by three complex numbers
$\epsilon_{e\mu},~\epsilon_{e \tau},~\epsilon_{\mu\tau}$,
defined in Eq.\eq{TreeLevel}. This gives rise to several relations between non-SM effects
in neutrino observables and in lepton flavour-violating processes.
Effects in charged leptons are suppressed by a one loop factor with respect to
effects in neutrinos:
\begin{equation}
P(\nu_i\to \nu_j; L\approx 0) \sim |\epsilon_{ij}^2|,\qquad
{\rm BR}(\ell_i\to \ell_j \gamma) \sim |e\epsilon_{ij}/4\pi|^2, \qquad
(i\neq j).
\end{equation}
Present bounds from neutrino experiments  and
from charged lepton processes are summarized in Table~1.
Due to the loop factor,
detectable neutrino effects are compatible with lepton flavour violating bounds,
unlike what is obtained with a generic larger set of SU(2)$_L$-invariant dimension six 
operators,
where neutrino and charged lepton effects both arise at tree level.
Values of $\epsilon_{ij}\circa{>}10^{-4}$ (including possible CP-violating phases)
will be probed by future neutrino experiments.
The importance of $\mu\to e\bar{e}e$ and $\mu\to e$ conversion relative to $\mu\to e \gamma$
can be enhanced, with respect to the usual 
magnetic-penguin dominance approximation,
if the right-handed neutrino is strongly coupled or 
if the cut-off of the theory is significantly larger than the $W$ mass.

All these effects can be generated in four dimensions by adding ``right-handed neutrinos''
with TeV-scale masses and order one Yukawa couplings,
but such choice of parameters is not motivated by neutrino masses.
On the contrary, extra-dimensional models that try to address the hierarchy problem 
and to generate neutrino masses give an
order-of-magnitude expectation for the $\epsilon_{ij}$ parameters of 
$\epsilon_{ij}\sim \ell_\delta (v/\TeV )^2\sim 10^{-{\rm few}}$ where $\ell_\delta$ is 
a loop factor in $\delta$ dimensions. Therefore, these models are compatible
with a ``natural'' value for the fundamental scale.

In the flat minimal model, the six $\epsilon_{ij}$ parameters are
related and can be expressed in terms of a single unknown $\epsilon$,
see Eq.~(\ref{eq:minimal}). The value of $\epsilon$ is at present constrained
by LEP and neutrino experiments ($\epsilon \circa{<} 0.004$) and by 
the $\mu \to e \gamma$ decay, see Eqs.~(\ref{lim1}) and (\ref{lim2}).
Improvements in the sensitivity on rare muon processes and measurements
at a future neutrino factory will significantly extend the probe on 
the hypothesis of an extra-dimensional origin of neutrino masses.

\paragraph{Acknowledgments}
We thank Paolo Cr{\bf i}minelli, Riccardo Rattazzi, Serguey Petcov and Alexei Smirnov
for useful discussions.

\paragraph{Note added}
The NuTeV collaboration~\cite{NuTeV} has recently reported
a 3-$\sigma$ anomaly in $\nu_\mu$-nucleon scattering, that can interpreted
as due to a ratio between the $\bar{\nu}_\mu Z \nu_\mu$
and the
$\bar{\nu}_\mu W \mu$ couplings $(0.58\pm 0.21)\%$ lower than in the SM.
 Since LEP found that the $Z$ couplings to other fermions
agree with SM predictions at the {\em per-mille} level, the NuTeV anomaly
could be due to physics beyond the SM that mainly affects the neutrinos.
This is what happens in the extra dimensional models considered here,
where the $\bar{\nu}_\mu Z \nu_\mu$ coupling is $1-\epsilon_{\mu\mu}$
smaller than in the SM, while the $\bar{\nu}_\mu W \mu$ coupling is
smaller by only $1-{\epsilon_{\mu\mu}}/{2}$ (see section 3). 
The NuTeV anomaly could be fitted by $\epsilon_{\mu\mu}=0.0116\pm 0.0042$.
This value gives a reduction of the $Z\to\nu\bar{\nu}$ width compatible
with the LEP measurement, 2-$\sigma$ lower than its SM prediction (see
footnote in page 7). However, the fact that the charged current is also
modified imposes severe constraints 
(the strongest bound, $\epsilon_{\mu\mu}+\epsilon_{ee}\circa{<}0.002$, comes from 
a comparision of the muon lifetime with precision electroweak data), 
which prevent a clean explaination the NuTeV anomaly.
% 
% such that a global fit to
% all data (including the new NuTeV anomaly) assuming there are
% extradimensional neutrinos would be only marginally better
% than the standard model fit.
% 

\appendix

\setcounter{footnote}{0}
\setcounter{equation}{0}
\section{Integrating out right-handed neutrinos}

In this Appendix, we present in detail the expressions for the effective
operators that mediate rare muon processes and the muon anomalous magnetic moment
when the entire tower of KK right-handed neutrinos (up to some arbitrary
cut-off) is integrated out.
The computation is done in the minimal flat model, but can be easily generalized to other cases of interest.

\subsubsection*{Magnetic moment type operator for $\mu \rightarrow e \gamma$}
The rare decay $\mu \rightarrow e \gamma$ takes place at the one-loop level.
The following magnetic moment operator is generated when all one-loop 
diagrams involving KK neutrinos are added.  
\begin{equation}
{\cal L} = m_\mu \bar{e} \sigma_{\alpha \beta} 
(A_L P_L +A_R P_R) \mu F^{\alpha \beta} +{\rm h.c.}
\end{equation}
where $A_L=0$,
\begin{eqnarray}
\label{func_meg}
A_R &=& \frac{eg^2}{64 \pi^2} \frac{1}{m^2_W} \sum_{\vec{n},j} V_{e \vec{n}}^j 
V_{\mu \vec{n}}^{j*}
G_\gamma (m^{(j)2}_{\vec{n}}/m^2_W),\\
G_\gamma(x) &=& -\frac{x(1-6x+3x^2+2x^3-6x^2\ln x)}{4(1-x)^4}.
\label{g_gamma}
\end{eqnarray}
Here the summation of all KK modes, which are labeled
by the generation $j$ and the integer vector $\vec{n}$ in the $\delta$-extra
dimension, are taken into account.
$m^{(j)}_{\vec{n}}$ is the mass of a KK neutrino labeled by 
$j$ and ${\vec{n}}$,
and $V_{i \vec{n}}^j$ is defined as $V_{i {\vec{n}}}^j\equiv 
U_{ij} V_L^{j{\vec{n}}}$, where $U_{ij}$
is the standard active neutrino mixing matrix and 
$V_L^{j \vec{n}}$ is the active-KK mixing matrix.
When $m^{(j)}_{\vec{n}} \gg m_{\nu j}$,
$m^{(j)2}_{\vec{n}} \approx {\vec{n}}^2/R^2$ and  
$V_L^{j \vec{n}} \approx m_{\nu j}/m^{(j)}_{\vec{n}}$.
The branching ratio for $\mu \rightarrow e \gamma$ is given by
\begin{eqnarray}
\frac{{\rm BR}(\mu \rightarrow e \gamma)}
{{\rm BR}(\mu \rightarrow e \nu_{e}\nu_{\mu})} &=& 
\frac{48 \pi^2 |A_R|^2}{G_F^2}=
 \frac{3 \alpha}{2\pi} \left|\sum_{\vec{n},j} V_{e \vec{n}}^j 
V^{j*}_{\mu \vec{n}}  G_\gamma(m^{(j)2}_{\vec{n}}/m^2_W) \right|^2.
\end{eqnarray}
The same expressions can be used for $\tau\rightarrow l\gamma$
after replacing $m_{\mu}\rightarrow m_{\tau}$, $V^{j*}_{\mu {\vec{n}}}
\rightarrow V^{j*}_{\tau {\vec{n}}}$, $V^{j}_{e {\vec{n}}}
\rightarrow V^{j}_{l {\vec{n}}}$, and 
${\rm BR}(\mu \rightarrow e \nu_{e}\nu_{\mu})\rightarrow
{\rm BR}(\tau \rightarrow l \nu_{l}\nu_{\tau})$, for $l=e$ or $\mu$.

Next, we sum over all KK modes. In models with large extra dimensions, 
the sum over the KK states can be accurately replaced by an integral
\begin{eqnarray}
\sum_{\vec{n}}^{|\vec{n}|< \Lambda R} 
f\left(\frac{{\vec{n}}^2}{{R}^2}\right) \rightarrow
S_\delta R^\delta \int_0^{\Lambda} dE \,E^{\delta-1} f(E^2),
\end{eqnarray}
where $f$ is any function and $S_\delta=2\pi^{\delta/2}/\Gamma(\delta/2)$. 
The cut off $\Lambda$ is expected to be of order of the fundamental scale 
$\Lambda \sim \bar{M}_D$. Therefore,
\begin{eqnarray}
\sum_{\vec{n} ,j} V_{e \vec{n}}^j V_{\mu \vec{n}}^{j*} 
G_\gamma(m_{\vec{n}}^{(j)2}/m^2_W)
%    &=& \sum_j U_{ej} U_{\mu j}^* 
%    \sum_{M=-\infty}^{\infty} |(V_L^j)_{1M}|^2
%    G_\gamma(m_M^{(j)2}/m^2_W) \nonumber \\
&\approx& \sum_j U_{ej} U_{\mu j}^* 
S_\delta R^\delta \int_0^{\Lambda} 
dE \, E^{\delta-1}\left| \frac{m_{\nu j}}{E} \right|^2
G_\gamma(E^2/m^2_W)
\nonumber
\\
&\approx& \sum_j U_{ej} U_{\mu j}^* m_{\nu j}^2
S_\delta R^\delta \int_0^{\Lambda} dE \, E^{\delta-3} \left(-\frac{1}{2}
\right)\nonumber \\
%   &=& -\frac{S_\delta}{2(\delta-2)} 
%   \left(\frac{\Lambda}{\bar{M}_D} \right)^\delta \frac{(2\pi R 
%   \bar{M}_D)^\delta}{(2\pi)^{\delta}}\frac{1}{\Lambda^2} 
%\sum_j U_{ej} U_{\mu j}^* m^2_{\nu j}
%   \nonumber \\
&=& -\frac{1}{2} \left[\frac{S_\delta}{\delta-2}
\left(\frac{\Lambda}{\bar{M}_D}\right)^\delta \left(
\frac{\bar{M}_4}{\bar{M}_D}\right)^{2\delta/\gamma}
\frac{1}{\Lambda^2}\frac{1}{(2\pi)^{\delta}}\right]
\sum_j U_{ej} U_{\mu j}^* m_{\nu j}^2.
\label{sum_meg}
\end{eqnarray}
We only consider $\delta>2$.
Note that we have used the definition of the reduced Planck mass, Eq.~(\ref{Mp}). 
The term in the square brackets equals $\epsilon/\Delta m^2_{\atm}$ (as defined in Sec.~2).

\subsubsection*{4-fermion operators for $\mu \rightarrow e\bar{e}e$ and $\mu\rightarrow e$
conversion in nuclei }
In addition to the magnetic moment type operator Eq.~(\ref{func_meg}),
the following 4-fermion operator contributes to the $\mu \rightarrow e\bar{e}e$
process:
\begin{eqnarray}
\cal{L} &=& \bar{e} \gamma_\alpha P_L \mu ~~
\bar{e} \gamma^\alpha \left\{
\left( A_\gamma + g^e_L A_Z +A_B \right)P_L+
\left( A_\gamma + g^e_R A_Z \right) P_R \right\} e +{\rm h.c.} 
\end{eqnarray}
Here $g^f_{L(R)}=T^f_{3L(R)}-Q_f \sin^2\theta_W$
($g^e_L=-1/2+\sin^2\theta_W$ and $g^e_R=\sin^2\theta_W$).
The coefficients $A_\gamma$, $A_Z$, and $A_B$ correspond to contributions from
photon penguin, $Z$-penguin, and box diagrams, respectively. Explicitly,
\begin{eqnarray}
A_\gamma &=& -\frac{e^2 g^2}{32 \pi^2} \frac{1}{m^2_W}
\sum_{\vec{n},j} V_{e \vec{n}}^j V^{j*}_{\mu \vec{n}} 
F_\gamma (m^{(j)2}_{\vec{n}}/m^2_W), \\
A_Z &=& -\frac{g^4}{64 \pi^2} \frac{1}{m^2_W}
\bigg\{ \sum_{\vec{n},j} V_{e \vec{n}}^j V_{\mu \vec{n}}^{j*}
 F_Z (m^{(j)2}_{\vec{n}}/m^2_W) \nonumber \\
&&
+\sum_{\vec{n},\vec{m},j} V_{e \vec{n}}^j V_L^{j \vec{n} *} 
V_L^{j \vec{m} *} V_{\mu \vec{m}}^{j*} G_Z(m^{(j)2}_{\vec{n}}/m^2_W,
m^{(j)2}_{\vec{m}}/m^2_W) 
\bigg\}, \\
A_B &=& -\frac{g^4}{64 \pi^2}\frac{1}{m^2_W}
\sum_{\vec{n},\vec{m},i,j} V_{e \vec{n}}^i V_{\mu \vec{n}}^{i*} 
V_{e \vec{m}}^j V_{e \vec{m}}^{j*}
G_B(m^{(i)2}_{\vec{n}}/m^2_W,m^{(j)2}_{\vec{m}}/m^2_W), \\
%\riga{where}\\
F_\gamma (x)&=&-\frac{x \left\{12-11x-8x^2+7x^3+2x(x^2-10x+12)\ln{x}\right\}}{12 
(1-x)^4},\\
F_Z(x) &=& \frac{5x(1-x+x\ln{x})}{(1-x)^2}, \\ 
G_Z(x,y)&=&\frac{1}{x-y}\left\{
\frac{x^2(1-y)\ln{x}}{1-x}-\frac{y^2(1-x)\ln{y}}{1-y}\right\}, \\
G_B(x,y)&=&\frac{1}{x-y}
\left[
\left(1+\frac{xy}{4}\right)
\left\{\frac{1}{1-x}+\frac{x^2 \ln{x}}{(1-x)^2}
-\frac{1}{1-y}-\frac{y^2 \ln{y}}{(1-y)^2} \right\}
\right. \nonumber \\
&&\left. -2xy \left\{ \frac{1}{1-x}
+\frac{x \ln{x}}{(1-x)^2}
-\frac{1}{1-y}-\frac{y \ln{y}}{(1-y)^2}
\right\} \right].
\end{eqnarray}
The branching ratio for $\mu \rightarrow e\bar{e}e$ process is
\begin{eqnarray}
{\rm BR}(\mu \rightarrow e\bar{e}e)
&=& \frac{1}{8 G_F^2} \bigg[
\left|A_\gamma+g_R^e A_Z\right|^2
+2 \left|A_\gamma+g_L^e A_Z+A_B\right|^2
+32\left|e A_R\right|^2 \left(\ln\frac{m^2_\mu}{m^2_e}
-\frac{11}{4} \right)
\nonumber \\
&&+ 4e A_R 
\left\{3A_\gamma^* +(g_R^e+2g_L^e)A_Z^* +2A_B^* \right\}
+{\rm h.c.}
\bigg].
\end{eqnarray}
In addition to the magnetic moment type operator Eq.~(\ref{func_meg}),
the following 4-fermion operator contributes to $\mu \rightarrow e$
conversion in nuclei:
\begin{eqnarray}
\cal{L} &=& \sum_{q=u,d} \bar{q} \gamma_\alpha q~~
\bar{e} \gamma^\alpha \left
(-Q_q A_\gamma +\frac{g^q_L+g^q_R}{2} A_Z
+B_B^q \right) P_L \mu,
 \\
B^d_B &=& -\frac{g^4}{128\pi^2}\frac{1}{m^2_W}
\sum_{\vec{n},i,j} V_{e \vec{n}}^j V_{\mu \vec{n}}^{j*} 
\left|V_{u_i d}\right|^2
G_B(m^{(j)2}_{\vec{n}}/m^2_W,m^2_{u_i}/m^2_W),
 \\
B^u_B &=& -\frac{g^4}{32 \pi^2} \frac{1}{m^2_W}
\sum_{\vec{n},j} V_{e \vec{n}}^j V_{\mu \vec{n}}^{j*} 
G_B(m^{(j)2}_{\vec{n}}/m^2_W,0).
\end{eqnarray}
The $\mu \rightarrow e$ conversion rate is 
\begin{eqnarray}
{\rm R}(\mu \rightarrow e)
&=& \frac{\alpha^3}{4\pi^2} \frac{Z_{\rm eff}^4}{Z}
\frac{|F(-m^2_\mu)|^2 m^5_\mu}
{\Gamma_{\rm \mu-capt}}
\bigg| 
(2Z+N)\left( \frac{g^u_L+g^u_R}{2}A_Z+B^u_B \right)
 \nonumber \\ 
&&+(Z+2N) \left( \frac{g^d_L+g^d_R}{2} A_Z 
+B^d_B \right)
-Z (2e A_R +A_\gamma)
\bigg|^2,
\end{eqnarray}
where $\Gamma_{\rm \mu-capt}$ is the muon capture rate in 
nuclei of interest~{\cite{suzuki}}, $Z$ and $N$ are the proton and
neutron numbers, respectively, $F(-m_\mu^2)$ is the nuclear
form factor and $Z_{\rm eff}$ is the nuclear effective charge~\cite{chiang}.
Numerically, these nuclear parameters are
$\Gamma_{\rm \mu-capt}=1.7 \times 10^{-18}~{\rm GeV}$~$(4.6 \times 10^{-19}~{\rm GeV})$, 
$Z_{\rm eff}=17.6$~$(11.6)$, and 
$|F(-m^2_\mu)|=0.535$~$(0.64)$ for 
$^{48}_{22}{\rm Ti}$~($^{27}_{13}{\rm Al}$)~\cite{suzuki,chiang}.
%
%\begin{eqnarray}
%&\Gamma_{\rm \mu-capt}=1.7 \times 10^{-18}~{\rm GeV},
%Z=22,Z_{\rm eff}=17.6,~|F(-m^2_\mu)|=0.535,~{\rm for~^{48}_{22} Ti,~and} \\
%\end{eqnarray}
%for $^{48}_{22}{\rm Ti}$ and 
%
%\begin{eqnarray}
%&\Gamma_{\rm \mu-capt}=4.6 \times 10^{-19}~{\rm GeV},
%Z=13,Z_{\rm eff}=11.6,~|F(-m^2_\mu)|=0.64,~{\rm for~^{27}_{13}Al}.
%\end{eqnarray}
%for $^{27}_{13}{\rm Al}$.

Summing over the KK states following the steps previously outlined, we obtain
\begin{eqnarray}
A_\gamma &\approx& -\frac{e^2 g^2}{64 \pi^2} \frac{1}{m^2_W}
\epsilon \sum_j U_{ej} U_{\mu j}^* \frac{m^2_{\nu j}}{\Delta m^2_{\atm}}
\left\{
\frac{7}{6} +\frac{1}{3} \left(
\ln\frac{\Lambda^2}{m_W^2}-\frac{2}{\delta-2}
\right) \right\},
\\
A_Z &\approx& -\frac{g^4}{64 \pi^2} \frac{1}{m_W^2}
\epsilon \sum_j U_{ej} U_{\mu j}^* \frac{m^2_{\nu j}}{\Delta m^2_{\atm}}
\left\{ 
\epsilon J_\delta  \frac{m_{\nu j}^2\Lambda^2}
{\Delta m^2_{\atm}m^2_W}\right. \nonumber \\
&&\left. +\left(3-2 \epsilon \frac{m^2_{\nu j}}{\Delta m^2_{\atm}} \right)
\left( \ln\frac{\Lambda^2}{m_W^2}-\frac{2}{\delta-2} \right)
-5
\right\},
\\
A_B &\approx& -\frac{g^4}{64 \pi^2}\frac{1}{m^2_W}
\epsilon \sum_j U_{ej} U_{\mu j}^* \frac{m^2_{\nu j}}{\Delta m^2_{\atm}}
\left\{
\epsilon J_\delta  \sum_i |U_{ei}|^2 \frac{m^2_{\nu i}\Lambda^2}
{\Delta m^2_{\rm atm}m_W^2}
\right.\nonumber \\
&&\left.+\epsilon \sum_i |U_{ei}|^2 \frac{m^2_{\nu i}}{\Delta m^2_{\atm}}
-1 \right\},
\\
B_B^d &=& -\frac{g^4}{64 \pi^2} \frac{1}{m^2_W}
\epsilon \sum_j U_{ej} U_{\mu j}^* \frac{m^2_{\nu j}}{\Delta m^2_{\atm}}
\frac{|V_{td}|^2}{8} \frac{m^2_{\rm top}}{m^2_W}
\left(
\ln \frac{\Lambda^2}{m^2_W} -\frac{2}{\delta-2} \right),
\\
B_B^u &\sim& ~{\rm small},
\\
e A_R &\approx& -\frac{e^2 g^2}{64 \pi^2} \frac{1}{m_W^2}
\frac{\epsilon}{2} \sum_j U_{ej} U_{\mu j}^* \frac{m_{\nu j}^2}{\Delta m^2_{\atm}}. 
\end{eqnarray}
where $J_\delta \equiv (1-\delta/2)^2\int_0^1\int_0^1 (zw)^{\delta/2-1}\ln(z/w)/(z-w)\,dz\,dw$, which is the number of order 1. 
Numerically, $J_3=0.23$, $J_4=0.43$, $J_5=0.55$, and  
$J_6=0.63$.

%%%   In particula $I_6 = 4^2(\pi^2/60 - 1/8)$, $I_5 = 3^2(\pi^2/16-5/9)$.
Note that, unlike the $\mu\rightarrow e\gamma$ case, 
most of the amplitudes depend not only on the neutrino mass-squared difference
(which is directly measured by neutrino oscillation experiments),
but also on the magnitude of the neutrino mass-squared.

\subsubsection{Magnetic moment operator for muon $g-2$}
The anomalous magnetic moment ($\delta a_{\mu}$) of the muon is defined as the coefficient 
of the effective operator  
\begin{eqnarray}
{\cal L} &=& \frac{e}{4 m_{\mu}} \delta a_\mu ~\bar{\mu} 
\sigma_{\alpha \beta} \mu F^{\alpha \beta}.
\end{eqnarray}
The contribution to muon anomalous magnetic moment induced by massive KK modes 
is given by  
\begin{eqnarray}
\delta a_\mu &=& \frac{g^2}{16 \pi^2} \frac{m^2_\mu}{m^2_W}
\bigg\{\sum_{\vec{n},j} \left|V_{\mu \vec{n}}^j 
\right|^2 f({m^{(j)2}_{\vec{n}}/m^2_W})
-\frac{5}{6}\bigg\}, \\
f(x) &=& \frac{10-43x+78x^2-49x^3+4x^4+18x^3\ln x}{12(1-x)^4}.
\end{eqnarray}
Note that we have subtracted the SM ($W,\nu$)-loop in order to calculate the 
new physics contribution. It is useful to separate the sum into the ``massless'' part
(we neglect the effect of the small active neutrino masses)
and the ``massive'' part. Using the unitarity of $ V_{\mu M}^j$ and the fact that 
$f(0)=5/6$, we can rewrite
\begin{eqnarray}
\delta a_\mu &=& \frac{g^2}{16 \pi^2} \frac{m^2_\mu}{m^2_W}
\sum_{\vec{n},j} \left|V_{\mu \vec{n}}^j \right|^2 
G_\gamma (m^{(j)2}_{\vec{n}}/m^2_W),
\end{eqnarray}
Here, the function $G_\gamma(x)\equiv f(x)-5/6$ is defined by Eq.~(\ref{g_gamma}).
Summing over the KK states (see Eq.~(\ref{sum_meg})) we obtain Eq.\eq{g-2}.
%   \begin{eqnarray}
%   \delta a_\mu &=& -\frac{g^2}{32 \pi^2} \frac{m^2_\mu}{m^2_W}\epsilon
%   \sum_{j} \left|U_{\mu j} \right|^2 \frac{m^2_{\nu j}}{\Delta m^2_{\atm}}.
%   \end{eqnarray}
%   
%   Note that, since $G_{\gamma} <0$, $\delta a_\mu$ is always negative. 

\footnotesize
\begin{multicols}{2}

\end{multicols}

\end{document}